\documentclass[preprint]{aastex62}
\usepackage{color, soul}
\usepackage{graphics}
\usepackage{subfigure}
\usepackage{cases}
\usepackage{gensymb}
\usepackage{changepage}
\usepackage{chngcntr}
\usepackage{combelow}

\begin{document}

\title{Galactic Center IRS13E: Colliding Stellar Winds or an Intermediate Mass Black Hole?}

\author[0000-0001-8812-8284]{Zhenlin Zhu}
\affil{School of Astronomy and Space Science, Nanjing University, Nanjing 210023, China}
\affil{Key Laboratory of Modern Astronomy and Astrophysics (Nanjing University), Ministry of Education, Nanjing 210023, China}
\affil{SRON Netherlands Institute for Space Research, Sorbonnelaan 2, 3584 CA Utrecht, The Netherlands}
\affil{Leiden Observatory, Leiden University, Niels Bohrweg 2, 2300 RA Leiden, The Netherlands}

\author[	
0000-0003-0355-6437]{Zhiyuan Li}
\affil{School of Astronomy and Space Science, Nanjing University, Nanjing 210023, China}
\affil{Key Laboratory of Modern Astronomy and Astrophysics (Nanjing University), Ministry of Education, Nanjing 210023, China}

\author[0000-0001-5800-3093]{Anna Ciurlo}
\affil{Department of Physics and Astronomy, University of California, Los Angeles, CA 90095, USA}

\author[0000-0002-6753-2066]{Mark R. Morris}
\affil{Department of Physics and Astronomy, University of California, Los Angeles, CA 90095, USA}

\author[0000-0001-8261-3254]{Mengfei Zhang}
\affil{School of Astronomy and Space Science, Nanjing University, Nanjing 210023, China}
\affil{Key Laboratory of Modern Astronomy and Astrophysics (Nanjing University), Ministry of Education, Nanjing 210023, China}

\author[0000-0001-9554-6062]{Tuan Do}
\affil{Department of Physics and Astronomy, University of California, Los Angeles, CA 90095, USA}

\author[0000-0003-3230-5055]{Andrea M. Ghez}
\affil{Department of Physics and Astronomy, University of California, Los Angeles, CA 90095, USA}

\email{zhuzl@smail.nju.edu.cn; lizy@nju.edu.cn}

\begin{abstract}
A small cluster of massive stars residing in the Galactic center, collectively known as IRS13E, is of special interest due to its close proximity to the central supermassive black hole Sgr~A* and the possibility that an embedded intermediate-mass black hole (IMBH) binds its member stars. 
It has been suggested that colliding winds from two member stars, both classified as Wolf-Rayet type, are responsible for the observed X-ray, infrared and radio emission from IRS13E.
We have conducted an in-depth study of the X-ray spatial, temporal and spectral properties of IRS13E, based on 5.6 Ms of ultra-deep {\it Chandra} observations obtained over 20 years.  These X-ray observations show no significant evidence for source variability.  
We have also explored the kinematics of the cluster members, using {\it Keck} near-infrared imaging and spectroscopic data on a 14-yr baseline that considerably improve the accuracy of stars' proper motions.  The observations are interpreted using 3-dimensional hydrodynamical simulations of colliding winds tailored to match the physical conditions of IRS13E, leading us to conclude that the observed X-ray spectrum and morphology can be well explained by the colliding wind scenario, in the meantime offering no support for the presence of a putative IMBH. 
An IMBH more massive than a few $10^3{\rm~M_\odot}$ is also strongly disfavored by the stellar kinematics. 

\end{abstract}

\section{Introduction}
\label{sec:intro}
The innermost parsec of the Galactic center (GC) hosts the nearest super-massive black hole (SMBH), commonly known as Sgr~A*,
as well as a population of more than 100 young, massive stars (\citealp{Maness2007,Bartko2010,Pfuhl2011,Do2013}), offering a unique laboratory to study star formation and evolution in the vicinity of a SMBH.
Most of these stars belong to a clockwise orbiting disk-like structure (\citealp{Genzel2003,Paumard2006, Lu2009, Bartko2009, Yelda2014}).
In addition, at just $\sim$3$\farcs$5 southwest of Sgr~A* lies a small star cluster\footnote{Here we follow the convention to call IRS13E a {\it cluster}, although it is not clear that it is a gravitationally bound system, and its size does not compare to the much more massive Nuclear Star Cluster, nor with the two other well-known young star clusters, Quintuplet and Arches, in the Galactic Center.} known as IRS13E \citep{Maillard2004}, which comprises several compact objects within a projected radius of $\sim$0$\farcs$3, including at least two Wolf-Rayet (WR) stars (a WN8-type and a WC9-type, designated as IRS13 E2 and E4, respectively) and one OB supergiant (designated E1), classified based on their near-infrared (NIR) spectra \citep{Paumard2006, Martins2007}. 
The clustering of these stars is evidenced by their similar westward proper motion, 
yet the estimated enclosed stellar mass is insufficient to hold together the cluster against the strong tidal field of Sgr~A*, provided that the actual distance of IRS13E from Sgr~A* is not much larger than the projected offset. 
To solve this issue, an intermediate-mass black hole (IMBH) with a mass of $\sim10^3{\rm~M_\odot}$ was proposed to reside in IRS13E and help prevent tidal disruption of the cluster (\citealp{Maillard2004}).
This IMBH scenario has made IRS13E a particularly interesting case, but the very existence of the putative IMBH remains inconclusive (\citealp{Schodel2005,Paumard2006,Fritz2010}).\\

Multi-wavelength observations have revealed rich information about IRS13E.
Early {\it Chandra} observation found an X-ray counterpart for IRS13E, which was interpreted as thermal emission induced by the colliding winds from a long-period binary of massive stars among the cluster members (\citealp{Coker2000,Coker2002}).
Deeper {\it Chandra} observations confirmed the thermal X-ray spectrum of IRS13E, which, when fitted with an optically-thin plasma model, showed a temperature of $\sim$2.0 keV and an unabsorbed 2--10 keV luminosity of $\sim$$2.0\times10^{33}{\rm~erg~s^{-1}}$ \citep{Wang2006}.  
Diffraction-limited NIR imaging observations afforded by the Very Large Telescope (VLT) have resolved IRS13E into a hierarchy of 19 discrete components (\citealp{Fritz2010}). 
In particular, source E3, which is located about midway between E2 and E4 and now resolved into six components within an extent of $\sim$0\farcs2, shows in its NIR spectrum no sign of absorption lines or broad emission lines that are characteristic of massive, windy stars.  
Instead, the spectrum and morphology of E3 suggest a blob of warm dust and ionized gas, likely caused by colliding winds from the two WR stars, E2 and E4.
With Very Large Array (VLA) multi-epoch observations at 1.3 cm, \citet{Zhao2009} found that different components of IRS13E have proper motions pointing in various directions, suggesting that local stellar winds might play a significant role in the motion of the ionized gas.
\\

Very recently, Atacama Large Millimeter/submillimeter Array (ALMA) observations have provided an increasingly high-resolution view of IRS13E in the millimeter continuum and the hydrogen recombination line H30$\alpha$ (\citealp{Tsuboi2017, Tsuboi2019}). 
At an angular resolution of $\sim$0\farcs03, the continuum distribution of the IRS13E complex shows general similarity with that seen in the NIR.
In particular, E3 appears extended and clumpy, dominating the 230--340 GHz continuum which is consistent with free-free emission from a $\sim$$10^4$~K ionized gas. 
The distribution and line-of-sight velocity of the H30$\alpha$ line around E3 were interpreted as tracing a tilted, rotating ring. 
The compact ring size and substantial rotation velocity imply an embedded IMBH with an inferred mass $M_{\rm BH} \approx 2.4 \times 10^{4}{\rm~M_{\odot}}$ (\citealp{Tsuboi2019}).
While Tsuboi et al. recognized the X-ray counterpart of IRS13E and took it as evidence for an accreting IMBH, the possibility that the X-ray emission is due to colliding winds (hence unrelated to a presumed IMBH) was not considered. 
Indeed, in various hydrodynamic simulations of the mutual interaction of winds from $\sim$30 WR stars (including E2 and E4) within the central parsec of the GC, IRS13E is consistently predicted to be a prominent X-ray source (\citealp{Cuadra2006, Cuadra2008, Russell2017, Ressler2018}).
More recently, combining {\it Chandra} and ALMA observations with hydrodynamic simulation results, \citet{Wang2020} reinforced the view that the X-ray emission from IRS13E is due to colliding winds between E2 and E4. 
\\

The recent advent of multi-wavelength observations of IRS13E thus 
motivates further investigation.
In this work, we explore the nature of IRS13E from the X-ray perspective, based on imaging-spectroscopic data afforded by archival {\it Chandra} observations with an ultradeep exposure of $\sim$5.6 Ms accumulated over eighteen years. 
Further assisted with kinematic measurements in the NIR band enabled by the Keck telescope, as well as numerical simulations tailored to model the colliding winds within IRS13E, we critically examine both the colliding wind and IMBH scenarios for IRS13E.
The remainder of this paper is organized as follows. We describe the {\it Chandra} and Keck observations in Section \ref{sec:data}.
In Section \ref{sec:multi}, the X-ray morphology of IRS13E, which consists of a compact core and an interesting, previously unknown tail-like feature\footnote{ \citet{Wang2020} independently identifies an extended X-ray morphology of IRS13E based on a similar set of {\it Chandra} data. Our interpretation for this morphology is different from theirs, as will be explained in later Sections.}, is presented in a close comparison with that seen in the NIR and radio bands. 
In Section \ref{sec:pm}, we report the kinematics (proper motion, line-of-sight velocity and velocity dispersion) of the main members of IRS13E. 
The X-ray temporal and spectral properties of IRS13E are analyzed in Section~\ref{sec:flux} and Section~\ref{sec:spec}, respectively. 
In Section~\ref{sec:sim} we present hydrodynamic simulation of colliding winds tailored to the physical condition of IRS13E and compare the predicted X-ray emission with observations. 
A discussion of the results and a summary are given in Section~\ref{sec:dis} and Section~\ref{sec:sum}, respectively.
Throughout this work, we adopt 8.0 kpc as the distance to the GC (e.g., \citealp{Gravity2019,Do2019sci}) and quote errors at 68\% confidence level, unless otherwise stated. \\

\section{X-ray and Near-Infrared Data}
\label{sec:data}
\subsection{\it{Chandra} Observations}
The inner few parsecs of the GC have been frequently observed by the {\it Chandra X-ray Observatory} since launch, chiefly with its Advanced CCD Imaging Spectrometer (ACIS).
In this work, we utilize the same dataset used in our previous study on the Sgr~A* jet candidate G359.944-0.052 (\citealp{Zhu2019}), which includes 47 ACIS-I observations taken between 1999 September and 2011 March, 38 ACIS-S observations with the High Energy Transmission Grating (HETG) taken between 2012 February and October, and 39 ACIS-S non-grating observations taken between 2013 May and 2017 July.
For the ACIS-S/HETG observations, we include only the zeroth-order image data, i.e., X-rays directly captured by the detector without any dispersion by the grating system. 
More information about the {\it Chandra} dataset can be found in \citet[Table 1 therein]{Zhu2019}.
\\

The data reduction procedure is the same as in \citet{Zhu2019}.
Briefly, we uniformly reprocessed the level-1 events files with CIAO v4.9 and calibration files CALDB v4.7.7, following the standard pipeline\footnote{http://cxc.harvard.edu/ciao}.
The CIAO tool {\it reproject\_aspect} was employed to calibrate the relative astrometry among the individual observations, by matching the centroids of commonly detected point sources (not including Sgr~A* and IRS13E).
The relative astrometry thus achieved has an accuracy of $\lesssim0\farcs$1.
We examined the light curve of each ObsID and removed time intervals contaminated by significant particle flares.
The resultant $\sim$5.58 Ms cleaned exposure, of which 1.42 Ms is from ACIS-I, 2.83 Ms from ACIS-S/HETG zeroth-order (hereafter HETG for brevity) and 1.33 Ms is from ACIS-S, allow for a significantly improved spectral quality and extended temporal baseline compared to previous X-ray studies of IRS13E (\citealp{Coker2002,Wang2006}).

\subsection{Keck Observations}
NIR observations of the central half parsec of the GC have been frequently conducted with the Near Infrared Camera II (NIRC2; PI: K. Matthews) and the integral-field OH-Suppressing Infra-Red Imaging Spectrograph (OSIRIS; \citealp{Larkin2006}) on the Keck II telescope.
Both the spectroscopic and imaging observations were obtained with the laser-guide-star adaptive optics (LGS AO) system, ensuring an angular resolution close to the diffraction limit.
In this work, we make use of an OSIRIS observation towards GC Southwest on May 26, 2009, which was taken in the Kn3 band (wavelength range 2.121--2.229 $\micron$) with a platescale of 35 mas.
Detailed information about this observation can be found in \citet{Do2013}. 
For the proper motion measurements described in Section \ref{sec:pm}, we utilize time-dependent positional information derived as a byproduct of a variability analysis of NIR sources in the GC \citep{Gautam2019}, based on NIRC2 Kp-band (effective wavelength 2.124 $\micron$) imaging observations taken between June 2005 and August 2018, in a total of 37 epochs.
Detailed descriptions on the reduction and analysis of the NIRC2 data can be found in \citet{Gautam2019}.

\section{Morphology}
\label{sec:multi}
\begin{figure}[ht]
     \centering
     \includegraphics[width=\textwidth]{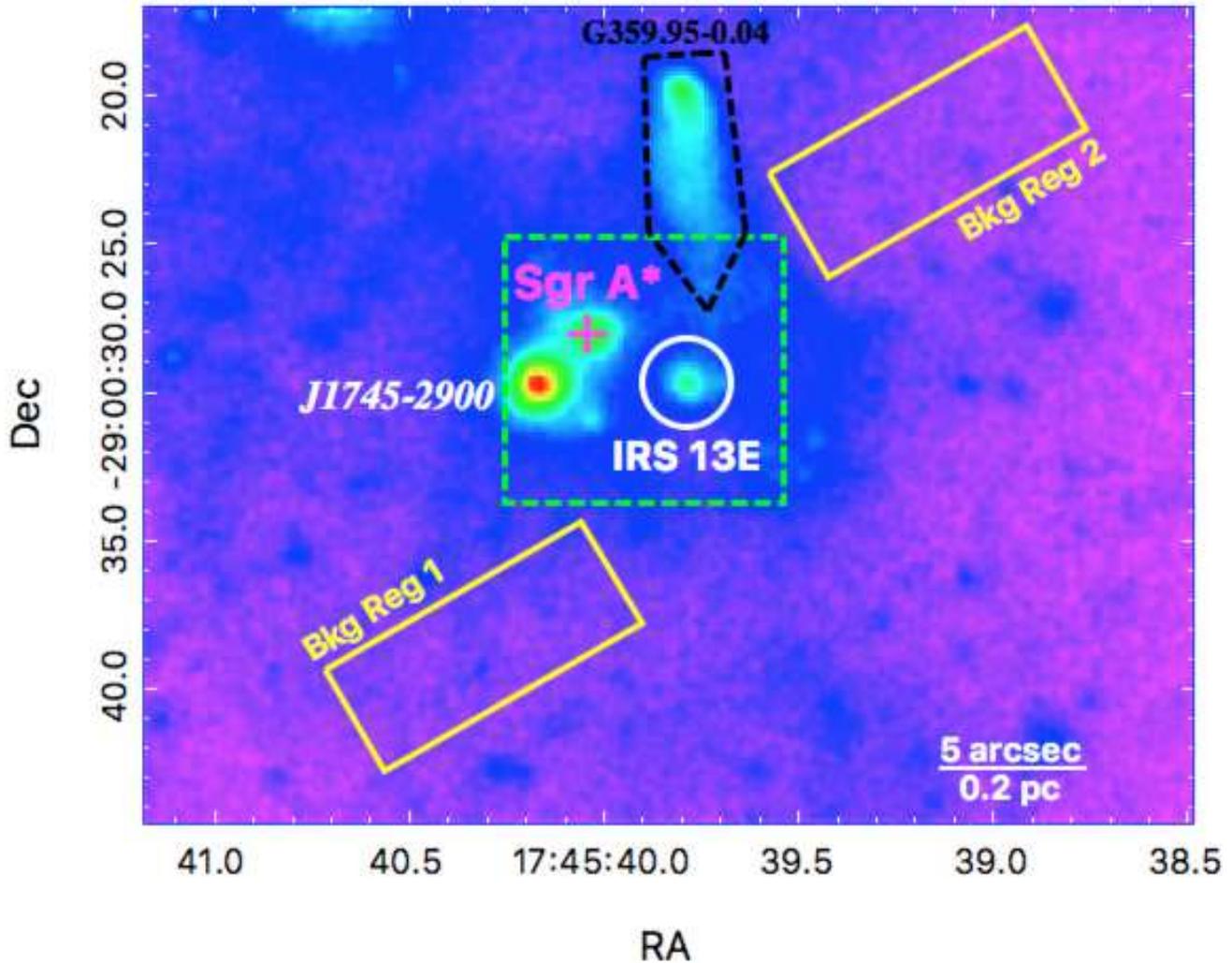} 
     \caption{A {\it Chandra}  2--8 keV image of the innermost region of the Galactic center, which combines 122 observations taken with ACIS-I, ACIS-S non-grating, and ACIS-S/HETG in zeroth-order. The image is smoothed with a Gaussian kernel of 2 pixels ($\sim$1\arcsec). Red (purple) color represents highest (lowest) intensities. The position of Sgr~A* is marked with a ``+'' sign.  The magnetar PSR J1745-2900 and the pulsar wind nebula candidate G359.95-0.04 are also marked, the latter with a dashed polygon. The 1$\farcs$5-radius circle denotes the source extraction region for the IRS13E complex, and the two yellow rectangles ($10\arcsec\times4\arcsec $ for each) outline the background regions for spectral analysis. The green dashed box outlines the zoom-in region shown in Figure~\ref{fig:match}.}
     \label{fig:chandra}
   \end{figure}

Figure \ref{fig:chandra} displays a {\it Chandra} 2--8 keV image of the vicinity of Sgr~A*, combining the $\sim$5.6 Ms ACIS data. 
X-rays from the GC with a photon energy below $\sim$2 keV are completely obscured by the strong line-of-sight extinction, hence we will focus on the 2--8 keV energy band hereafter. 
IRS13E, marked by a white circle, is among the brightest X-ray sources in this region, the others including Sgr~A*, the magnetar PSR J1745-2900 (\citealp{Kennea2013}) and a candidate pulsar wind nebula, G359.95-0.04 (\citealp{Wang2006}). 
Before closely examining the X-ray properties of IRS13E, our first task is to clarify the spatial relation between this compact X-ray source and the spatially-resolved emission observed at longer wavelengths, e.g., NIR and millimeter, where the available angular resolution is typically much higher than that of {\it Chandra}.
This will be crucial for understanding the exact origin of the multi-wavelength emission from the IRS13E complex, E3 in particular.  
While reports of precise positions/centroids for individual cluster members are present in the literature, a multi-wavelength cross-match was often absent or sometimes mistaken. 
\\

In principle, one can align the X-ray and NIR (or radio) images by cross-matching commonly detected sources. 
However, the only available source pairs within the common field-of-view are Sgr~A* and IRS13E itself. 
Therefore, we try to measure the relative offset of IRS13E from Sgr~A*, taking the latter as a fixed reference point for all wavelengths.
We focus on the 38 HETG observations, which were taken within the same calendar year, during which the effect of intrinsic proper motion (see Section~\ref{sec:pm}) is minimal.
To determine the X-ray centroid of IRS13E, we include all 2.8 Ms of the HETG exposure, whereas for Sgr~A* we only include time intervals during which Sgr~A* experiences significant X-ray flares, thus effectively manifesting itself as a point-like source (\citealp{Wang2013}).
To identify the flares, we have applied a Bayesian Blocking algorithm (\citealp{Scargle2013}) using an approach similar to that described in \citet{Mossoux2017}. 
An up-to-date analysis of Sgr~A*'s X-ray flares detected by {\it Chandra} will be presented elsewhere (G. Witzel et al. in preparation). 
\\

\begin{figure}
     \centering
      \includegraphics[width=0.8\textwidth]{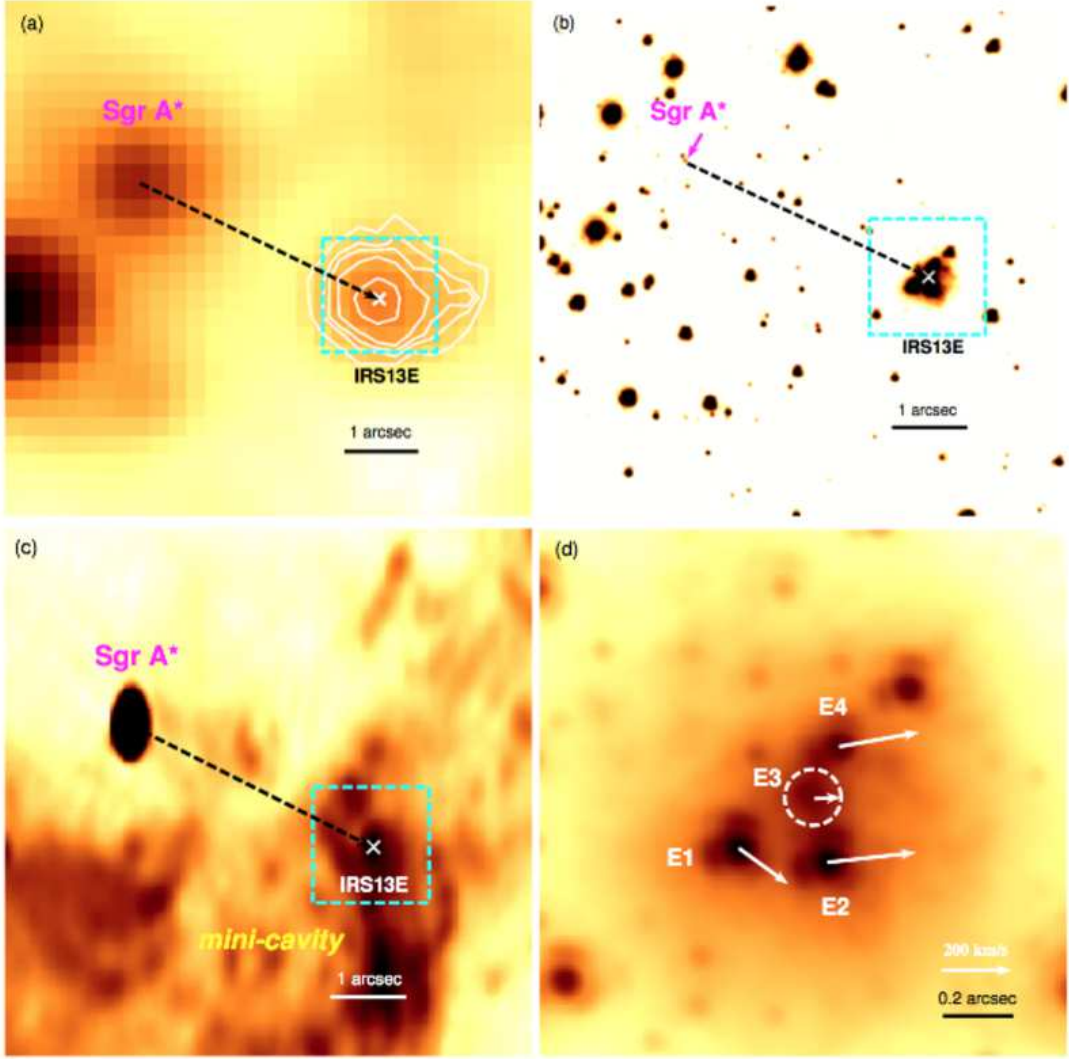}
     \caption{(a) {\it Chandra} 2--8 keV image of the region defined by the green dashed box in Figure \ref{fig:chandra}. The white intensity contours highlight the morphology of the X-ray source IRS13E, which exhibits a core plus a tail-like feature. 
The same region is shown in (b) Keck NIRC2 Kp-band image, taken on 2019 May 13 (\citealp{Do2019}), with the magenta arrow pointing to the position of Sgr~A*, and
(c) VLA X-band image (central frequency 8.6 GHz) taken on 2015 February 20, as part of project 14A-209 (PI: G.~Bower).  
In panels (a), (b) and (c), the white ``X''  marks the X-ray centroid of IRS13E, and the region enclosed by the cyan dashed box as seen by the NIRC2 Kp-band image is further zoomed in (d), where the proper motion of individual sources (E1, E2, E3 and E4) are marked by arrows rooting at the source centroid. The dashed circle indicates the aperture used for the kinematic analysis (Section~\ref{sec:pm}).
}
     \label{fig:match}
   \end{figure}

For each source, the centroid and its statistical error are determined using a maximum likelihood method (\citealp{Sarazin1980,Wang2004,Zhu2018}), iterating over the detected counts within a $0\farcs5$-radius circle. 
The measured centroid of IRS13E has a positional offset $(\Delta \alpha, \Delta \delta) = (-3\farcs25\pm0\farcs01, -1\farcs57\pm0\farcs01)$ from Sgr~A*, corresponding to a count-weighted epoch of 2012.8 (here and below the quoted offsets are measured in the ICRS coordinate system).
The rather small statistical uncertainty is due to the large number of counts detected for both sources ($\sim$1100 for Sgr~A* and $\sim$900 for IRS13E).
If a larger aperture of 1$\arcsec$-radius were adopted, the relative offset becomes $(\Delta \alpha, \Delta \delta) = (-3\farcs30\pm0\farcs01, -1\farcs61\pm0\farcs01)$, which might be due to the intrinsic extent of the source (see below).
This X-ray offset is to be contrasted with the offset of the NIR source E3, the centroid of which was reported to be ($-3\farcs$18, $-1\farcs$53) relative to Sgr~A* in the VLT Ks-band image at epoch 2005.4 \citep{Fritz2010}, and is also measured to be ($-3\farcs$20, $-1\farcs$53) in our Keck Kp-band image at epoch 2012.0 (see Section~\ref{sec:pm}).
Furthermore, a radio source has been identified as a probable counterpart of E3 from a VLA 7\,mm image taken at epoch 2011.6 showing an offset of $(-3\farcs20\pm0\farcs03, -1\farcs56\pm0\farcs03)$ \citep{Yusef2014}, 
and from a VLA 34.5 GHz image at epoch 2014.2, with an offset of $(-3\farcs21\pm0\farcs03, -1\farcs55\pm0\farcs03)$ \citep{Yusef2015}.
More recently, \citet{Tsuboi2017} reported an offset of $(-3\farcs19\pm0\farcs01, -1\farcs56\pm0\farcs01)$ from ALMA 340 GHz observations taken at epoch 2016.7. 
Figure \ref{fig:match} illustrates the multi-wavelength cross-matching of IRS13E.
From the above quoted offsets, we can conclude that (i) the NIR and radio centroids of E3 are highly coincident with each other, as previous work suggested, and (ii)
the X-ray centroid of IRS13E significantly deviates from the NIR/radio centroid of E3, by $\sim$$0\farcs05$ to the west, 
and even more so from either E2 or E4, being $\sim$$0\farcs15$ predominantely along the north-south direction (Figure \ref{fig:match}d). 
\\
     
\begin{figure}[h!]
     \centering
     \includegraphics[width=0.6\textwidth]{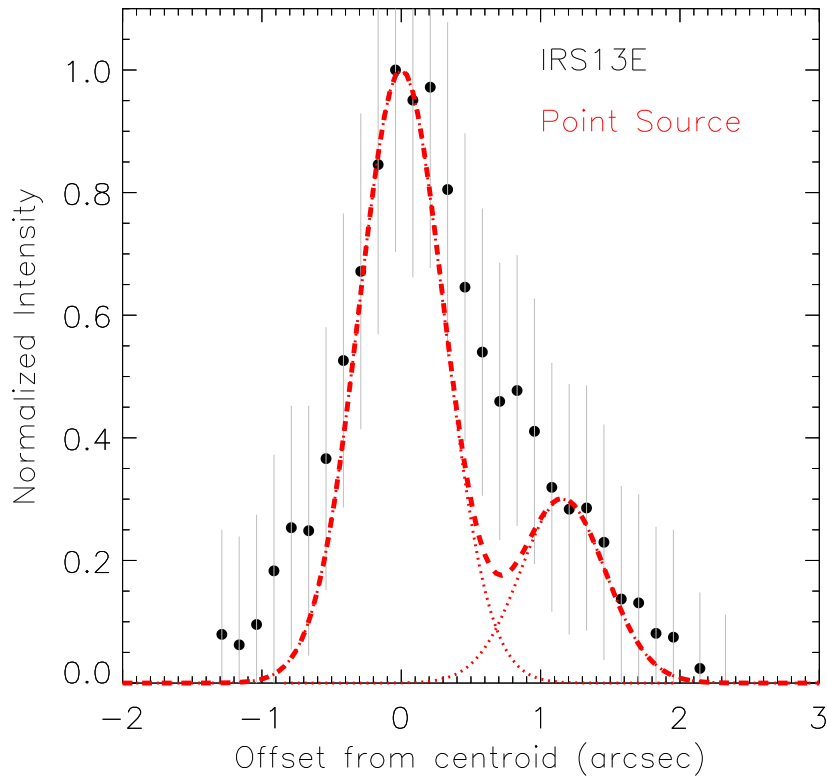}
     \caption{The one-dimensional X-ray intensity profile of the IRS13E complex along the east-west direction, showing a `tail'' extending to the west (positive offset). For comparison, the average profile of several nearby point sources, mimicking the local PSF, is shown as the dotted red curve. A PSF of lower intensity is placed at an offset of $1\farcs1$ to mimic the contribution from a possible second point source. The thick dashed curve is the sum of the two.
     }
       \label{fig:tail}
   \end{figure}

The deep {\it Chandra} image reveals that the compact X-ray source of IRS13E has a tail-like feature extending $\sim$1\farcs5 west from the source centroid, as highlighted by the white intensity contours in Figure \ref{fig:match}a. 
The orientation of this tail is approximately parallel with the proper motions of E2/E3/E4 (see Section \ref{sec:pm}).
In Figure \ref{fig:tail}, we plot the one-dimensional X-ray intensity profile of IRS13E, integrated along the east-west direction with a vertical full width of $2\farcs8$ and normalized to the peak value at the centroid (zero point). 
A local background running parallel to the source has been subtracted. 
We note that the immediate vicinity of IRS13E has a rather irregular, diffuse X-ray background and might be contaminated by the extended tail of G359.95-0.04 to the north, but our result here is unaffected thanks to the relative brightness of IRS13E.  
The intensity profile confirms the existence of the tail extending to the west (positive offset), when compared to the average intensity profile of several nearby X-ray point sources (\citealp{Zhu2018}), which is constructed in the same manner as for IRS13E and mimics the local point-spread function (PSF; shown as the red dotted curve). 
Moreover, apparently this tail cannot be fully accounted for by an additional point source, as illustrated by the second PSF in Figure \ref{fig:tail}.
We find that the fractional flux of this tail, estimated by subtracting the modeled PSF, is $\sim$32\% of the total.
We also compare the hardness ratio (HR) of the point-like core and the tail (for this purpose the latter is counted over an offset between $0\farcs8-1\farcs8$), which is defined as HR $\equiv$ (H-S)/(H+S) over the 2--4 keV ($S$) and 4--8 keV ($H$) bands. It turns out that the tail has HR $\approx 0.25\pm0.06$, significantly harder than the core with HR $\approx -0.07\pm0.03$.
The possible origin of this hard tail is further addressed in Section~\ref{sec:dis}.
\\




\section{Kinematic Analysis}
\label{sec:pm}
\begin{figure}
     \centering
    \includegraphics[width=0.95\textwidth]{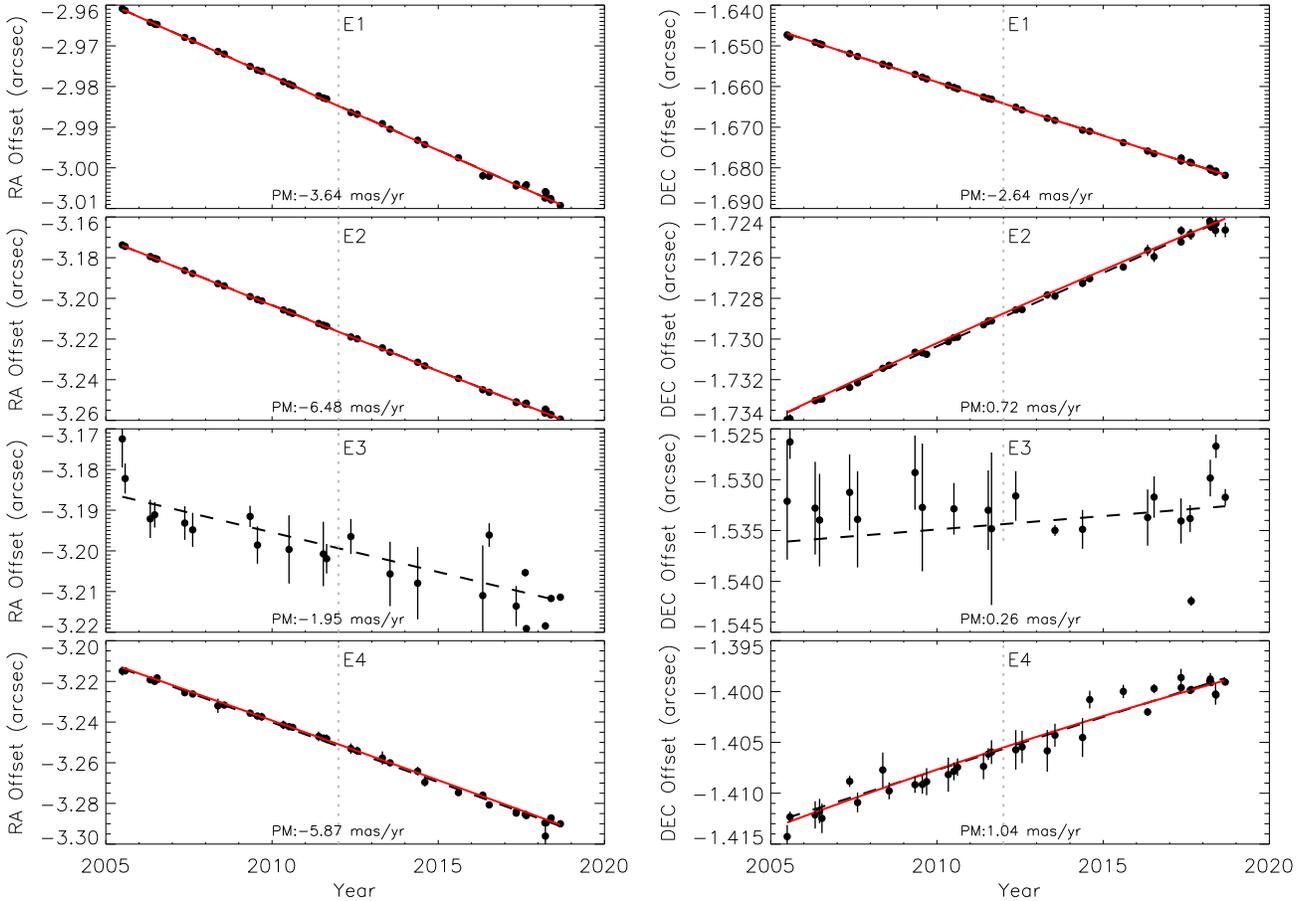}
     \caption{Proper motion measurements for the major components of the IRS13E complex. The {\it left} and {\it right} columns are for offsets along R.A. and Decl., respectively. The black dashed (red solid) lines show the best linear (parabolic) fit, with the fitted proper motion denoted in the individual panels. No parabolic fit is warranted for E3 due to the large fluctuations in its positional measurements. The vertical dashed line marks the epoch of 2012.0. 
     } 
\label{fig:fitpm}
   \end{figure}

The time-dependent positional offsets (in R.A. and Decl. with respect to Sgr~A*) of the four major components of IRS13E, namely, E1, E2, E3 and E4, are obtained with the same analysis as in \citet{Gautam2019} and \citet{Do2019} and are shown in Figure~\ref{fig:fitpm}.
These sources, in particular E2 and E4, show a similar westward motion, as first pointed out by \citet{Maillard2004}. 
It is apparent that E3 shows significant fluctuations in its position measurement, as compared to the other three sources. 
Already noticed in previous work \citep{Fritz2010}, this is an artifact due to the fact that E3 is an extended source containing a few sub-components. 
Since E4 locates in close proximity to E3, the Decl. measurements of E4 is also irregularly biased, due to contamination from the latter.
We characterize the time-dependent positions with a parabolic function (red solid lines in Figure \ref{fig:fitpm}), including the first-order (i.e., proper motion) and second-order (i.e., acceleration) terms,

\begin{subequations}
\begin{equation}
   \Delta\alpha = {\Delta\alpha}_0 + v_\alpha (t-t_0) + a_{\alpha} (t-t_0)^2, 
\end{equation}
\begin{equation}   
    \Delta\delta =  {\Delta\delta}_0 + v_\delta (t-t_0) + a_{\delta} (t-t0)^2,
\end{equation}
\end{subequations}
with R.A. and Decl. treated independently.
The derived offset (with respect to Sgr~A*), proper motion and acceleration of the individual objects at a nominal epoch $t_0$ = 2012.0 are summarized in Table \ref{tab:pm}.
Our measurements have smaller uncertainties than previous work based on VLT observations (\citealp{Paumard2006, Fritz2010, Eckart2013}), thanks to our longer temporal baseline.
We note that for E3 the second-order term is not fitted, since its positional fluctuations do not allow for a meaningful determination. 
Acceleration is constrained for the other three sources. A significant acceleration is only found in the case of E2 along R.A., at the $\sim$$4\,\sigma$ level.
The limits of $a_{\alpha}$ and $a_{\delta}$ allow us to constrain the mass of the putative IMBH ($M_{\rm BH}$), for the specific case in which an IMBH is embedded in E3. 
In this case, the acceleration follows

\begin{subequations}
\begin{equation}
a_{\alpha} = -0.8\frac{(M_{\rm BH}/10^4{\rm~M_\odot})x}{(D_{\rm GC}/{\rm 8~kpc})^3(R^2+z^2)^{\frac{3}{2}}}~\mu{\rm as~yr^{-2}}, 
\end{equation}
\begin{equation}
a_{\delta} = -0.8\frac{(M_{\rm BH}/10^4{\rm~M_\odot})y}{(D_{\rm GC}/{\rm 8~kpc})^3(R^2+z^2)^{\frac{3}{2}}}~\mu{\rm as~yr^{-2}}, 
\end{equation}
\end{subequations}
where $x$ and $y$ are the projected offsets (in arcseconds) from E3, $R^2=x^2+y^2$, and $z$ is the line-of-sight depth. 
Since we have no observational constraint for $z$, we assume for the moment that $z^2 = \frac{1}{2}R^2$. 
Using the centroid positions and 3$\sigma$ limits in $a_{\alpha}$ and $a_{\delta}$ (Figure~\ref{fig:fitpm}), 
this assumption leads to an upper limit on $M_{\rm BH}$ for each case of E1, E2 and E4.
The tightest constraint comes from $a_{\delta}$ of E2, which corresponds to $M_{\rm BH} < 2.2 \times10^3{\rm~M_\odot}$. 
Conversely, assuming $M_{\rm BH} \approx 2 \times10^4{\rm~M_\odot}$ as claimed by \citet{Tsuboi2019}, this would require $|z| > 2.3 R$.
A similar constraint is obtained from $a_{\alpha}$ of E1, while E4 provides no strong constraint due to its larger positional uncertainties. 
These limits only become slightly looser if the few percent uncertainty of the GC distance is taken into account \citep{Gravity2019,Do2019sci}.
Therefore, the stellar kinematics strongly disfavor the presence of an IMBH more massive than a few $10^3{\rm~M_\odot}$, unless all three stars have an elliptical orbit and the long-axis is close to the line-of-sight.  
We stress that the above arguments have neglected the gravity of Sgr~A*, which is valid only if the IRS13E cluster lies at a line-of-sight distance ${|z|}_{13E} \gtrsim0.4 (10^4{\rm~M_\odot}/M_{\rm BH})$ parsec from Sgr~A*. This condition may not be satisfied, for instance, if IRS13E coincides with the western rim of the mini-cavity (Figure~\ref{fig:match}c; \citealp{Zhao2009}, their Fig. 21).
\\

\begin{deluxetable}{l c c c c c c c c c r} 
\tabletypesize{\footnotesize}
\tablewidth{-20pt}
\tablecaption{Kinematics of the major components of IRS13E}
\tablehead{
\colhead{Object}&
\colhead{${\Delta}\alpha$}&
\colhead{${\Delta}\delta$}&
\colhead{$v_{\alpha}$}&
\colhead{$v_{\delta}$}&
\colhead{$\bar{v}_{\alpha}$}&
\colhead{$\bar{v}_{\delta}$}&
\colhead{$\bar{v}_{r}$}&
\colhead{$a_{\rm \alpha}$}&
\colhead{$a_{\rm \delta}$}
\\
\colhead{(1)} &
\colhead{(2)} &
\colhead{(3)} &
\colhead{(4)} &
\colhead{(5)} &
\colhead{(6)} &
\colhead{(7)} &
\colhead{(8)} &
\colhead{(9)} &
\colhead{(10)} 
}
\startdata
E1 &-2.985 & -1.664 &  -3.64$\pm$0.01   & -2.64$\pm$0.01     &-142.2$\pm$0.3 & -103.1$\pm$0.3&-9$\pm$8 & $2.16\pm1.82$&$2.11\pm1.91$\\
E2 &-3.216 & -1.729 &  -6.48$\pm$0.01   &  0.72$\pm$0.01   &-253.3$\pm$0.3 & 28.3$\pm$0.3 &-46$\pm$5&$8.44\pm2.00$&$-3.59\pm2.03$\\
E3 &-3.199 & -1.534 & -1.95$\pm$0.14    & 0.26$\pm$0.13   & -76.2$\pm$5.4 & 10.3$\pm$5.0 &-34$\pm$3 & --   &  --  \\
E4 &-3.253 & -1.407 &  -5.87$\pm$0.03   & 1.04$\pm$0.02    & -229.7$\pm$1.0 & 40.8$\pm$0.9 &71$\pm$6 &$-9.63\pm 10.39$&$-10.83\pm8.35$
\enddata
\tablecomments{(1) Object name. (2)-(3) R.A. and Decl. offsets relative to the position of Sgr~A* at epoch 2012.0, in units of arcsec. The positional uncertainties are on the order of 1 mas.  (4)-(5) R.A. and Decl. proper motions with respect to Sgr~A*, in units of mas~yr$^{-1}$. (6)-(7) R.A. and Decl. proper motions in units of km~s$^{-1}$, assuming a distance of 8.0 kpc. 
(8) LSR Line-of-sight velocity,  in units of km~s$^{-1}$. (9)-(10) R.A. and Decl. accelerations, in units of $\mu$as~yr$^{-2}$. 
}
\label{tab:pm}
\end{deluxetable}

We measure the radial velocity of the IRS13 sources by extracting each of their NIR spectra from the OSIRIS data. 
The applied aperture radius for E3 is $0\farcs08$, while for background extraction we utilize an annulus with radii between $0\farcs11$ and $0\farcs18$, masking E1, E2 and E4. 
The Kn3-band spectrum, covering from 2.121 to 2.229 $\micron$, mainly includes three emission lines: 2.1450/2.2184 $\micron$ [FeIII] lines and 2.1661 $\micron$ HI Br-$\gamma$ line. 
By measuring the Doppler shift of the Br-$\gamma$ line in each spectrum, we obtain the source velocities along the line-of-sight. 
These radial velocities are then corrected for solar motion with respect to the local standard of rest (LSR).
All results are summarized in Table~\ref{tab:pm}. 
Taking the three-dimensional velocities (proper motion plus line-of-sight) 
of E2 and E4 with respect to their mean velocity and assuming that the two stars are gravitationally bound by a point mass at their barycenter,
we estimate a value of $M_{\rm dyn} \gtrsim 2.6\times10^{3}~{\rm M}_{\odot}$. If E1 were also included, the estimated enclosed mass becomes $M_{\rm dyn} \gtrsim 6.9\times10^{3}~{\rm M}_{\odot}$.
These values are close to the above estimates based on proper motion. 
The line broadening can also act as a good tracer of the central enclosed mass, in the case that gas dynamics are gravitationally determined rather than dominated by winds or shocks.
We have measured the line widths of the three emission lines in the E3 spectrum, resulting in a one-dimensional velocity dispersion of 90$\pm$10 km~s$^{-1}$, which would correspond to an enclosed mass of $M_{\rm dyn} \approx 1.7\times10^{4}~{\rm M}_{\odot}$ within the spectral extraction radius of 0\farcs08 centered at E3 for an isotropic velocity field.  
This velocity dispersion, however, might be instead attributed to the interaction of stellar winds from E2 and E4 (see discussion in Section~\ref{sec:dis}).
We defer a detailed study of the spatially-resolved emission lines of the IRS13E complex to future work.
\\

\section{X-ray and Near-Infrared Flux Variability}
\label{sec:flux}
\begin{figure}[h!]
     \centering
      \includegraphics[width=0.7\textwidth]{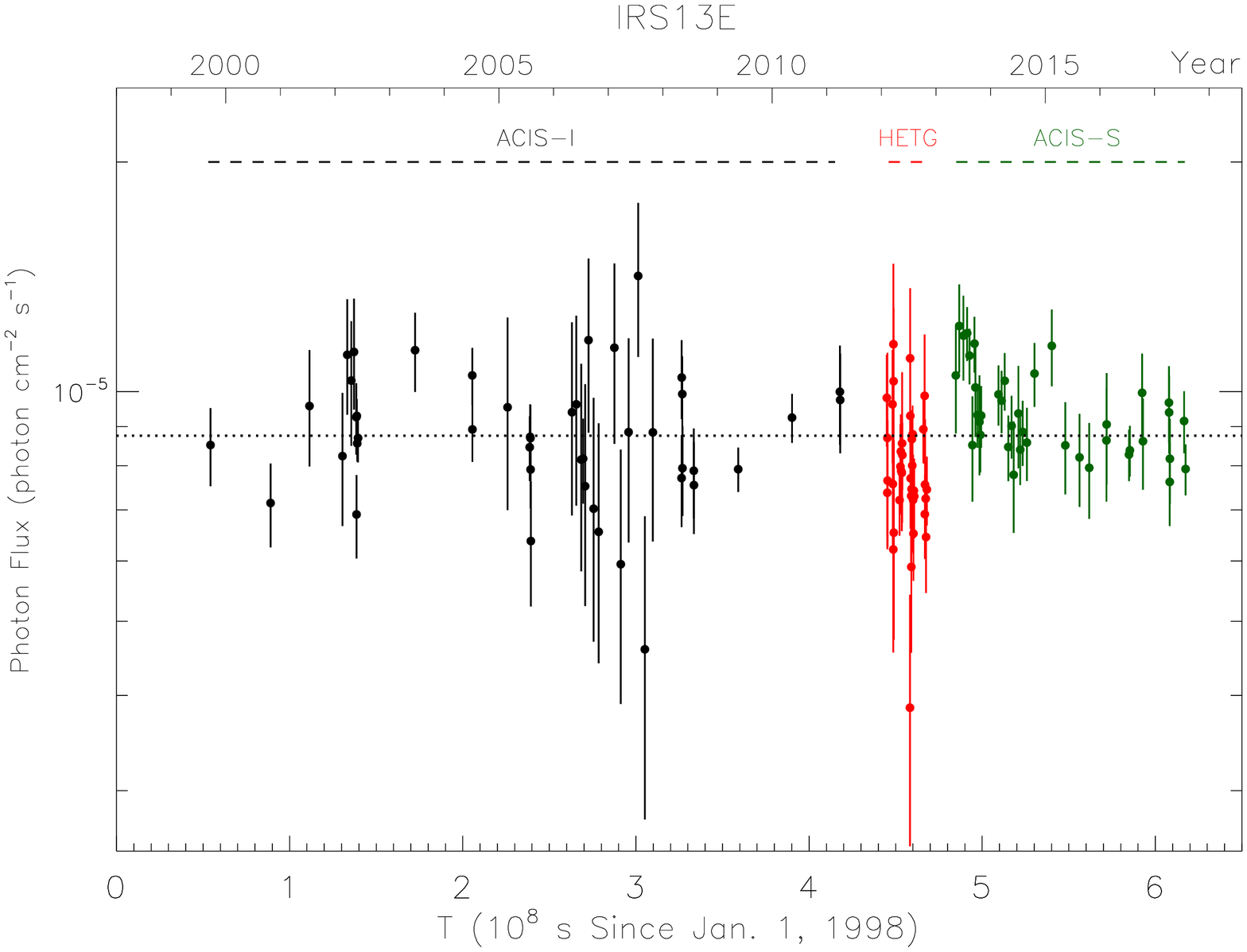}
      \includegraphics[width=0.7\textwidth]{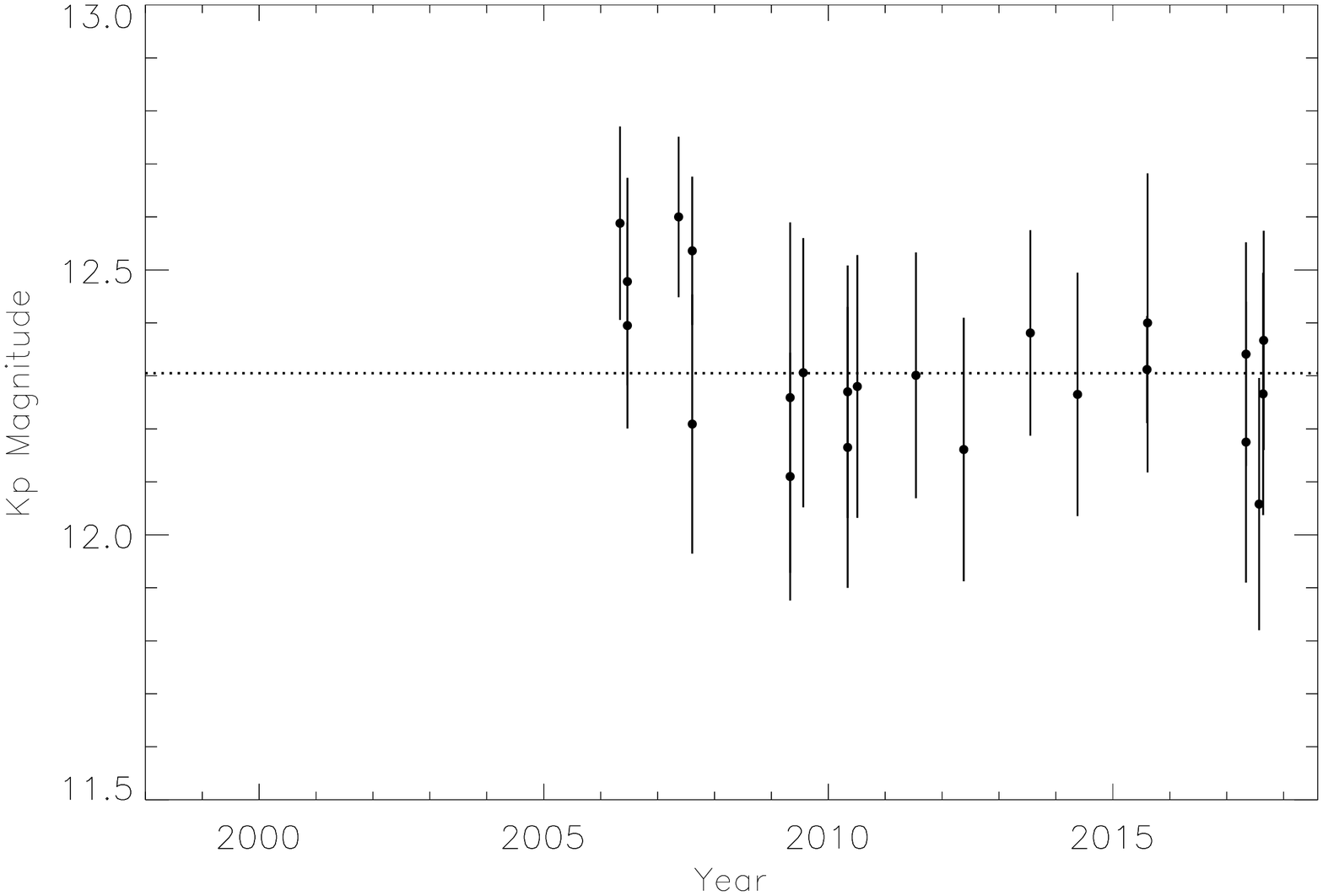}\\
     \caption{{\it Upper panel}: The 2--8 keV light curve of IRS13E, including all ACIS-I (black), HETG (red) and ACIS-S (green) data after correction for the effective exposure. {\it Lower panel}: The light curve of E3 measured in the Kp-band. In both panels, the error bars represent 1-$\sigma$ uncertainty, and the dotted lines denote the average flux.} 
     \label{fig:lc}
   \end{figure}

We now turn to examine the flux variation of IRS13E during the $\sim$18-year observing period in the X-ray band.
For X-ray flux measurements, the source region is defined by a $1\farcs5$-radius circle, which necessarily includes the ``tail", whereas for the background we use two nearby rectangular regions as illustrated in Figure \ref{fig:chandra}.
We utilize the CIAO tool {\it aprates} to compute the observed photon flux and bounds for each observation, corrected for the local effective exposure. 
The resultant long-term 2--8 keV light curve is displayed in the upper panel of Figure \ref{fig:lc}. 
IRS13E shows little flux variation overall; the most significant variation was seen in the period of May--Oct 2013 (green data points), during which a $\sim$30\% flux drop is observed at $\sim 1.8\,\sigma$ significance.
To further quantify the flux variability, we calculate the {\it normalized excess variance} following \citet{Nandra1997}, 
$\sigma_{\rm exc}^{2} \equiv \sum_{i=1}^N[(f_i-\overline{f})^2-\sigma_{f,i}^2]/(N\overline{f}^2) \approx 4.05\times 10^{-3}$, 
where $f_i$ and $\sigma_{f,i}$ are the measured flux and statistical error in the {\it i}-th observation, and $N=122$ is the total number of measurements. 
We find $\sigma_{\rm exc}^{2} = (4.05\pm4.76) \times 10^{-3}$, which indicates that the apparent small deviations from the mean photon flux, $\overline{f} \approx 8.7\times 10^{-6}~{\rm ph~cm^{-2}~s^{-1}}$, are predominantly statistical fluctuations rather than significant intrinsic variability. 
We further search for short-term X-ray variability in IRS13E, by employing the CIAO tool {\it glvary} on each of the 122 observations.
This tool uses the \citet{Gregory1992} variability test algorithm on the unbinned X-ray data.
The result strongly suggests that there is neither significant short-term (i.e., intra-observation) variability in IRS13E over the past two decades.
\\

We also examine the Kp-band light curve of E3, which is plotted in the lower panel of Figure \ref{fig:lc}.
The measured Kp-band magnitude of E3 has a relatively large uncertainty due to its extended shape.
We find an average Kp-band magnitude $m_{\rm Kp} = 12.3\pm$0.13, corresponding to a flux density of $85\pm10$ mJy.
The normalized excess variance of the NIR flux is calculated to be $\sigma_{\rm rms}^{2} = (-2.2\pm 0.6) \times 10^{-2}$, also indicating no significant intrinsic flux variation.
We note that \citet{Fritz2010} reported a flux increase of 0.25 mag in the H- and Ks-band between 2004--2006, which may be coincident with the first few data points in our Kp-band light curve. 
\\


\section{X-ray Spectral Analysis}
\label{sec:spec}
The X-ray spectra of IRS13E are extracted for each ObsID using the CIAO tool {\it specextract}, adopting the same source and background regions as described in Section~\ref{sec:flux}.
The tool also generates the corresponding ancillary response files (ARFs) and redistribution matrix files (RMFs). 
Next we produce a combined spectrum of ACIS-I, HETG or ACIS-S, by co-adding ObsIDs taken with the same instrument and weighting the ARFs and RMFs by the effective exposure. 
Distinguishing the three instruments is necessary since they have very different response over the energy range of interest, whereas for ObsIDs taken with the same instrument, the response varies little with time.  
\\

\begin{figure}[h!]
     \centering
     \includegraphics[angle=270,width=0.8\textwidth]{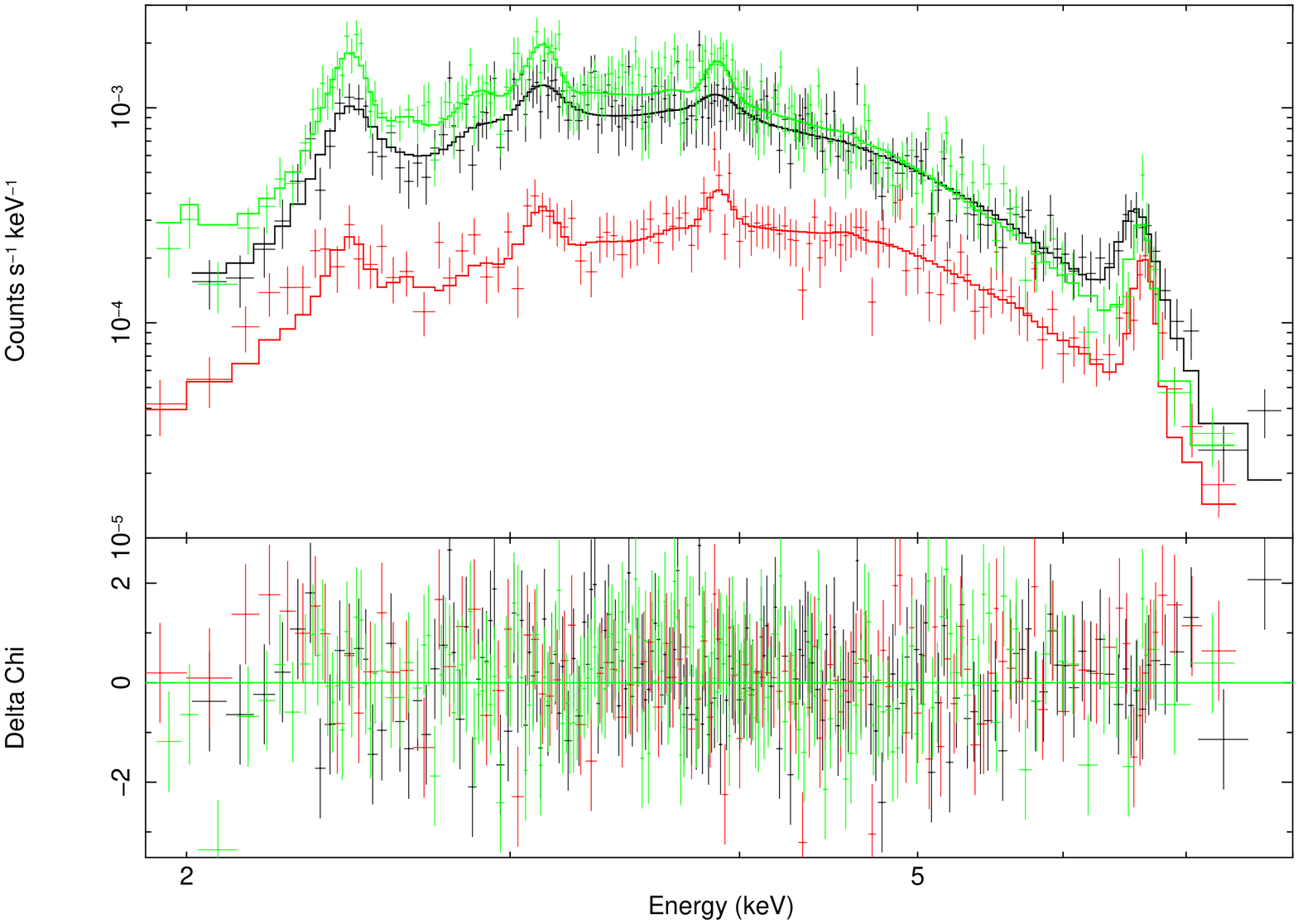}
     \caption{Co-added spectra of IRS13E, adaptively binned to achieve S/N $\geq 3$ per bin. The ACIS-I/HETG/ACIS-S spectrum is shown in black/red/green color. Also plotted are the best-fit absorbed {\it vnei} models.} 
     \label{fig:spec}
   \end{figure}

The three sets of spectra are shown in Figure \ref{fig:spec}, which shows that they have comparable signal-to-noise ratios (S/N).
Significant emission lines from S, Ar, Ca and Fe are present, strongly suggesting that the X-ray spectrum of IRS13E is dominated by thermal emission.
First we attempt to jointly fit the three spectra using an optically thin plasma model with line-of-sight absorption (model {\it tbabs*vapec} in XSPEC), assuming that the plasmas have reached collisional ionization equilibrium (CIE) and allowing for a varied abundance among S, Ar, Ca and Fe. The abundances of other elements which show no significant lines in the spectrum are fixed at unity. Here we adopt the elemental abundance standard of  \citet{wilms2000}.
All parameters except for the normalization are tied among the three spectra. 
In particular, the line-of-sight absorption toward the GC should not have significantly varied during the 18-year timespan.
This model results in an unsatisfactory fit with $\chi^{2}/dof = 549.9/484$, showing deviations mainly at the emission lines. 
This indicates an imbalance in the ionization level of different elements, which can be the case if the hot plasma in IRS13E is not in CIE. 
We therefore apply an absorbed non-equilibrium ionization (NEI) model ({\it tbabs*vnei} in XSPEC) to jointly fit the three spectra.
The NEI model is employed with the following free parameters: electron temperature ($kT$), density-weighted ionization timescale ($\tau$; in units of cm$^{-3}$ s), abundance of elements S, Ar, Ca, Fe, and normalization scaled with the volume emission measure. 
The abundances of individual elements are allowed to vary, but the value of a given element is kept tied among the three spectra. 
We initially allowed for a varied $kT$, but the consistent values found among the three spectra prompt us to also tie this parameter in the joint fit.
\\

The absorbed {\it vnei} model fits the spectra quite well, as illustrated in Figure \ref{fig:spec}.
The $\chi^{2}/dof$ has significantly improved to 489.4/484.
The fitted absorption column density, $N_{\rm H} = 16.65^{+0.39}_{-0.38} \times 10^{22}~{\rm cm}^{-2}$, is compatible with previous measurements of nearby X-ray sources (\citealp{Wang2013, Zhu2019}).
The best-fit gas temperature, $kT\sim$ 2.0 keV, demands a continuous energy input to prevent the hot gas from cooling down; furthermore, the low ionization timescale $\tau \sim (1-2) \times 10^{11}~{\rm cm^{-3}~s}$ supports the case for NEI, since $\tau > 10^{12}~{\rm cm^{-3}~s}$ is usually required for the plasma to reach ionization equilibrium \citep{2010ApJ...718..583S}.
The best-fit model predicts an absorption-corrected 2--10 keV luminosity of $L_{\rm 2-10} = 1.94^{+0.01}_{-0.06}\times 10^{33}$, $1.64^{+0.01}_{-0.07}\times 10^{33}$, $2.13^{+0.03}_{-0.06}\times 10^{33}$~erg~s$^{-1}$, for the ACIS-I, HETG and ACIS-S spectra, respectively.
Hence the spectral analysis reveals a $\sim$20\% intrinsic flux variation among the three spectra, which is buried in the statistical fluctuation when examining the photon fluxes of individual observations (Section~\ref{sec:flux}). 
The spectral fit results are summarized in Table~\ref{tab:spec}. 
\\

\begin{deluxetable}{lccr} 
\tabletypesize{\small}
\tablewidth{-50pt}
\tablecaption{X-ray spectral fit results}
\tablehead{
&
\colhead{ACIS-I}&
\colhead{HETG}&
\colhead{ACIS-S}
}
\startdata
$N_{\rm H}~(10^{22}~{\rm cm}^{-2})$ & $16.65^{+0.39}_{-0.38}$ & --  &  -- \\
$kT~({\rm keV}) $ & $1.92^{+0.09}_{-0.08}$ & --  &  --\\
S  &$1.07^{+0.11}_{-0.11}$ &--&-- \\
Ar &$1.25^{+0.20}_{-0.19}$ &--&--\\
Ca & $1.32^{+0.24}_{-0.23}$  &--&--\\
Fe & $0.70^{+0.08}_{-0.07}$ & -- & --\\
$\tau~(10^{11}~\rm cm^{-3}~\rm s)$  & $1.72^{+0.38}_{-0.48}$ &$2.21^{+0.47}_{-0.64}$&$1.16^{+0.19}_{-0.21}$\\ 
norm $(10^{-4}~\rm cm^{-5})$  & $5.46^{+0.53}_{-0.52} $& $4.81^{+0.47}_{-0.46} $ & $6.13^{+0.61}_{-0.60} $ \\
$L_{2-10}~(10^{33}~$erg~s$^{-1})$ & $1.85^{+0.02}_{-0.06}$ & $1.64^{+0.02}_{-0.07}$ & $2.07^{+0.03}_{-0.06}$ \\[5pt]
\hline
$\chi^{2}/\rm d.o.f$  & &489.43/484 \\[5pt]
\enddata
\tablecomments{For HETG and ACIS-S spectra, the column densities and the abundances of S, Ar, Ca and Fe (relative to the interstellar medium abundances of \citealp{wilms2000}) are tied to the values of ACIS-I spectra when carrying out the joint-fit. The quoted errors are at 1-$\sigma$ confidence level.}
\label{tab:spec}
\end{deluxetable}

\section{Numerical simulation of colliding winds}
\label{sec:sim}
The observed thermal X-ray spectrum of IRS13E is suggestive of shock-heated plasma in colliding stellar winds \citep{Coker2002}. 
In this section we perform numerical simulations to test this specific scenario, in which IRS13E is produced by colliding winds from the two WR stars, namely, E2 and E4. 
We emphasize that our goal here is to provide a quantitative comparison with the X-ray data, rather than a comprehensive modeling of the multi-wavelength observations.   
The simulations are carried out using a hydrodynamics (HD) code, \textit{PLUTO}\footnote{http://plutocode.ph.unito.it/} \citep{Mignone2007, Mignone2012}.
A 3-dimensional HD frame with a cartesian grid of 256$^3$ is set to construct a physical box of 0.08$^3$ pc$^3$, sufficient to enclose the IRS13E complex and in the meantime provide an equivalent angular resolution of 8 mas. 
To avoid accidental errors, the HLL (Harten, Lax, Van Leer) approximate Riemann Solver, a more robust solver, is applied in the simulation.
We assume an ideal equation of state (EoS) for gas and take into account radiative cooling, which is approximated by an implemented piecewise power-law function. 
\\

The primary parameters of the two-star system include the physical separation of the two stars ($d$), the wind mass loss rate ($\dot{M}_{\rm w}$) and the wind terminal velocity ($v_{\rm w}$) of each star. 
Since there is no observational constraint on the differential line-of-sight distance between the two stars, we assume that they lie in a plane perpendicular to the line-of-slight, hence their physical separation equals the physical separation, $d = 0.012$ pc (converted from the angular separation of $0\farcs32$ between E2 and E4 for a line-of-sight distance of 8 kpc). 
We further neglect the motion of the two stars. Our test simulations including the proper motions (Table~\ref{tab:pm}) find that this is a reasonable assumption, since the stars move slowly compared to their wind velocities. 
The mass loss rates and terminal velocities are empirically determined according to the NIR spectroscopic analysis of \citet{Martins2007}. 
However, our test simulations show that the predicted X-ray luminosity would be much larger than the observed values (Table~\ref{tab:spec}), when we adopt the original mass loss rates given by \citet{Martins2007}.
Therefore, we adopt a $\sim$2.5 times lower mass loss rate for both stars in our fiducial simulation, approximately scaled to match the observed X-ray luminosity of the system.
This lower mass loss rate ($1.8\times10^{-5}$ M$_{\odot}$ yr$^{-1}$) is likely still within the uncertainty allowed by the NIR spectral modeling and is not atypical of WR stars.
Moreover, spectroscopically-determined mass loss rates can often be overestimated due to wind clumping (e.g., \citealp{2016AdSpR..58..761R}). 
Radio observations of the WR and O-stars near SgrA* also indicate that their mass loss rates measured from NIR spectra can be overestimated by a factor of 3 to 10 (\citealp{Yusef2015}).
We note that a good match to the observed X-ray luminosity can still be obtained with a higher mass loss rate if a larger physical separation $d$ were adopted, due to the well-known scaling relation of $L \propto \dot{M}^2_{\rm w} d^{-1}$ in colliding winds in the adiabatic limit (\citealp{Stevens1992}). 
\\

The winds are injected at a radius of 0.002 pc from each star, with the wind terminal velocity of 750 km~s$^{-1}$ for E2 and 2200 km~s$^{-1}$ for E4 from \citet{Martins2007}. Effects of gravity and radiation by both stars on the wind velocity are neglected, given their large separation.
We also adopt an initial circumstellar medium (CSM) temperature of $10^4$ K and a CSM density of 10 cm$^{-3}$, a value appropriate for the diffuse hot gas in the central parsec of the GC (\citealp{Baganoff2003}). Our simulation results are insensitive to these choices, since the strong stellar winds will blow out the CSM in roughly the dynamical time ($\lesssim$40 yr). 
The simulation time is set to 200 years, which is sufficiently long to reach a quasi-steady configuration in the colliding wind region, but sufficiently short for the two stars to not significantly change their relative distance.
All the relevant parameters in our fiducial simulation are summarized in Table~\ref{table:parameters}.
\\

\begin{figure*}
    \centering
    \includegraphics[width=0.47\textwidth]{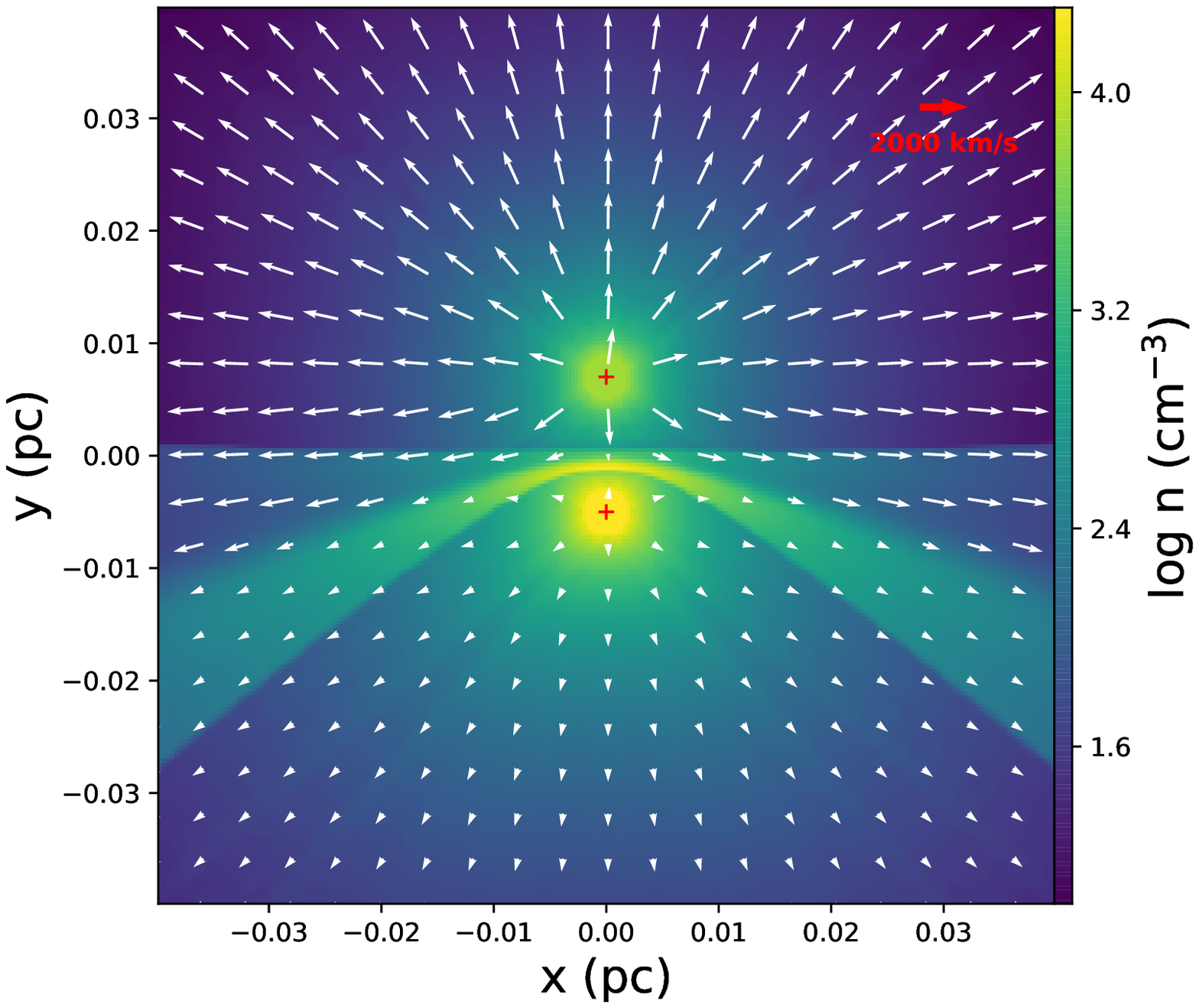}
    \includegraphics[width=0.49\textwidth]{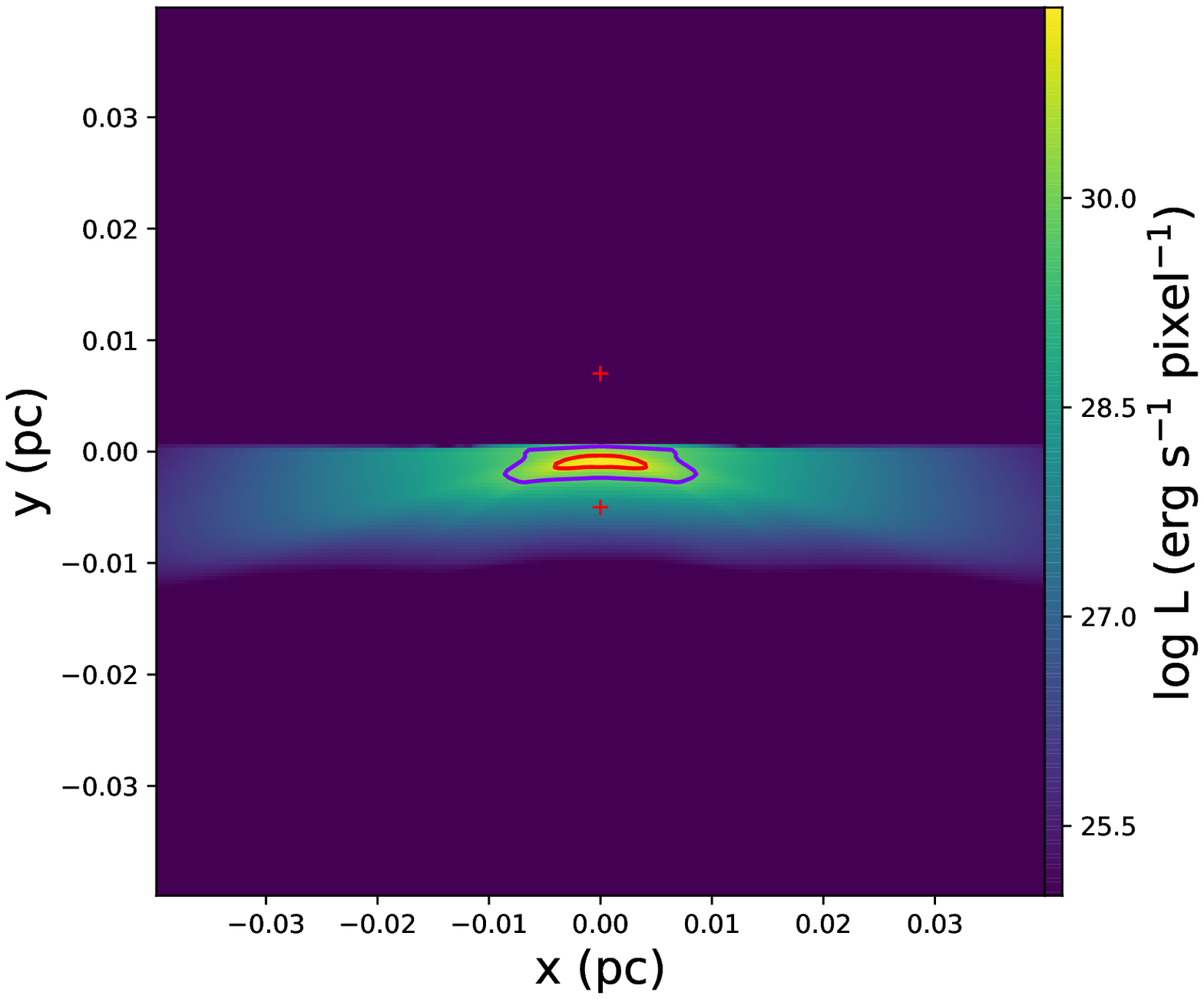}
    \caption{Illustration of the fiducial colliding wind simulation. {\it Left}: Density distribution in the $z = 0$ plane, overlaid by vectors representing the local velocity.     
    {\it Right}: 2--8 keV X-ray surface brightness distribution integrated along the $z$-axis. Significant X-ray emission is concentrated in the colliding wind region running roughly parallel to the x-axis.
    The red (blue) contour outlines the region enclosing 50\% (90\%) of the total X-ray emission. In both panels, positions of the two WR stars are marked by `+' signs.  
      }
\label{fig:simu}
\end{figure*}

\begin{figure*}
    \centering
      \includegraphics[angle=90,width=0.7\textwidth]{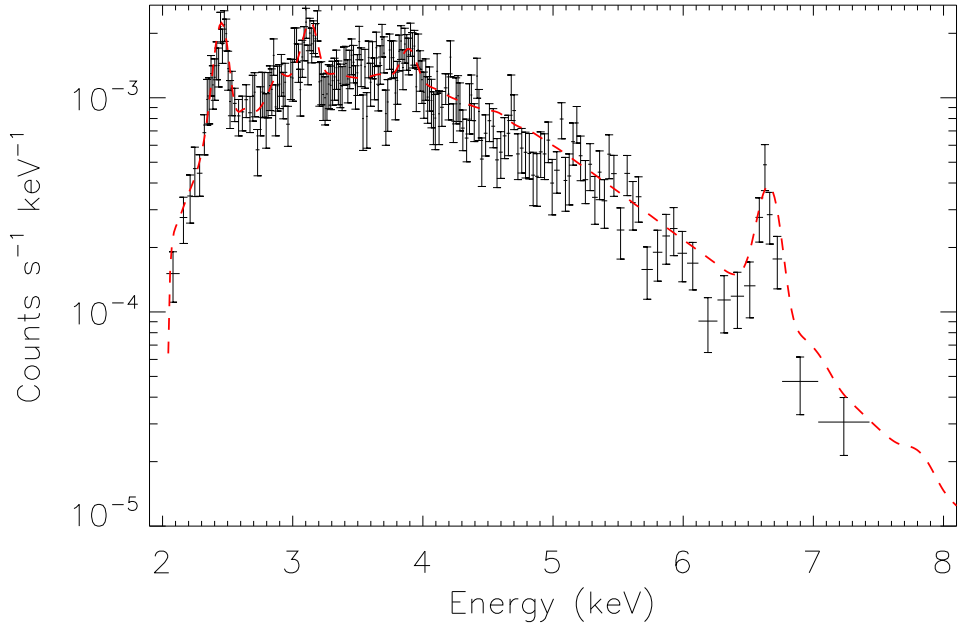}
    \caption{A comparison between the observed ACIS-S spectrum (black data points) and the simulation-predicted spectrum (red dashed curve). The latter has been multiplied by the energy-dependent foreground absorption and convolved with the instrumental response.} 
\label{fig:simu_spec}
\end{figure*}

The left panel of Figure~\ref{fig:simu} shows the particle density distribution in the $z = 0$ plane of the fiducial simulation. Here the $z$-axis is aligned with the line-of-sight. 
The two WR stars, E2 and E4, are placed at $(x, y, z) = (0, -0.005, 0)$ pc and $(x, y, z) = (0, 0.007, 0)$ pc, respectively. 
A narrow and curved colliding wind region (CWR) can be clearly seen between the two stars. The location and curvature of the CWR are expected given the $\sim$ 1:3 ratio of wind momentum ($\dot{M}_{\rm w} v_{\rm w}$) of the two stars (\citealp{Usov1992}). 
The particle temperatures in the CWR span a range of $10^{6.0}$--$10^{8.0}$ K, thus X-ray emission is expected from the CWR. 
We calculate the X-ray spectrum radiated by the $i$th pixel in the simulated box as $L_i=1.2 n_i^2 \Lambda(T_i, Z, E) V_i$, where $n_i$ is number density, $T_i$ is the temperature, $\Lambda$ is the volume emissivity as a function of $T_i$, abundance $Z$ and photon energy $E$, and $V_i$ the physical volume of the pixel (same for all pixels).
$\Lambda$ is extracted from ATOMDB\footnote{http://www.atomdb.org} version 3.0.9, for which we adopt the NEI model with an ionization timescale of $1.2\times10^{11}\rm~cm^{-3}~s$ and specific abundances of S, Ar, Ca and Fe, to be consistent with the best-fit spectral model (Table~\ref{tab:spec}).
Intrinsic absorption due to the wind particles is neglected since we are only concerned with the 2--8 keV energy range. 
The right panel of Figure~\ref{fig:simu} shows the X-ray surface brightness distribution of the fiducial simulation, integrated along the $z$-axis and over 2--8 keV.
It can be seen that 50\% of the X-ray emission (traced by the red contour) comes from a projected region of $\sim$0.01 pc $\times$ 0.002 pc  ($\sim0\farcs26\times0\farcs05$), which is consistent with the observed X-ray centroid position within the uncertainty (Section \ref{sec:multi}).
The simulated spectrum, multiplied by the foreground absorption and convolved with the instrumental response, is contrasted with the observed ACIS-S spectrum in Figure~\ref{fig:simu_spec}. The two spectra are in remarkably good agreement. 
The total 2--8 keV luminosity from the simulation is found to be $2.0\times10^{33}{\rm~erg~s^{-1}}$, which also matches well with the observed value (Table~\ref{tab:spec}).
\\

\begin{table}
  \caption{Summary of Simulation Parameters}
  \label{table:parameters}
  \centering
  \begin{tabular}{l l l}
      \hline\hline
      Two Star Parameters             &  IRS13E2               & IRS13E4\\
      \hline
      Mass Loss Rate (10$^{-5}$ M$_{\odot}$ yr$^{-1}$)   & 1.8  & 1.8 \\
      Wind Terminal Velocity (km s$^{-1}$)         & 750  & 2200 \\
      Wind Injection Radius (pc)          & 0.002           & 0.002 \\
      \hline
      \hline
      Other Parameters                & \\
      \hline
      Separation (pc)                    &    0.012 &  \\
      CSM Density (cm$^{-3}$)          &  10     & \\
      CSM Temperature (K)       &   $10^4$  &  \\
      Total Simulation Time (yr)                     & 200         & \\
      Mean Atomic Weight              & 1.3              & \\
      Adiabatic Coefficient           & 1.7            & \\
      \hline\hline
  \end{tabular}
\end{table}

\section{Discussion}
\label{sec:dis}

In the previous sections, we have shown a remarkable match between the observed thermal X-ray spectrum of IRS13E and the synthesized spectrum based on our fiducial simulation. 
Moreover,  the X-ray centroid is found at a position roughly midway between the two WR stars, again consistent with the simulations. 
We stress that the setup of the fiducial simulation is tightly constrained by the observed properties of the two WR stars, having few tunable parameters.
This lends strong support to the colliding wind scenario, in which copious X-rays are produced in the shocked-wind region. 
Nevertheless, not all observed properties of IRS13E are readily reproduced by our fiducial simulation. Below we shall address some of the most notable aspects.
\\ 

The observed X-ray morphology of IRS13E exhibits an appreciable extent, showing a tail-like feature containing $\sim$30\% of the total 2--8 keV flux (Section~\ref{sec:multi}), whereas in the simulation the projected X-ray-emitting region is highly concentrated and symmetric about the axis defined by the two stars (Figure~\ref{fig:simu}), which is a natural outcome of the symmetric winds. In principle, ram pressure due to the CSM may produce a tail-like structure. However, in this case the orientation of the tail should be due to a westward ram pressure, inconsistent with the mean proper motion of the two stars. A westward ram pressure, instead, may be due to a hypothetical outflow produced or collimated by Sgr~A* (e.g., \citealp{1996ApJ...460L..33M}).   
The strong winds of E2/E4 lead to a ram pressure on the order of ${\rho}_wv_w^2 \sim 10^{-5}$~g~cm$^{-1}$~s$^{-2}$ near the colliding wind region (Figure~\ref{fig:simu}). To balance this ram pressure, the hypothetical outflow from Sgr~A* would need to have an unrealistically large radial velocity of $10^4 {\rm~km~s^{-1}}$, given an empirical CSM number density of $\sim$10~cm$^{-3}$ \citep{Baganoff2003}. For a more realistic radial velocity of $\lesssim10^3 {\rm~km~s^{-1}}$, a bow shock will occur eastward of IRS13E, emitting X-rays which however are not seen. We have run a test simulation incorporating such an external flow to verify this picture, finding that an outflow from Sgr~A* cannot reproduce the observed X-ray morphology.  
\\

\begin{figure*}
    \centering
    \includegraphics[width=0.47\textwidth]{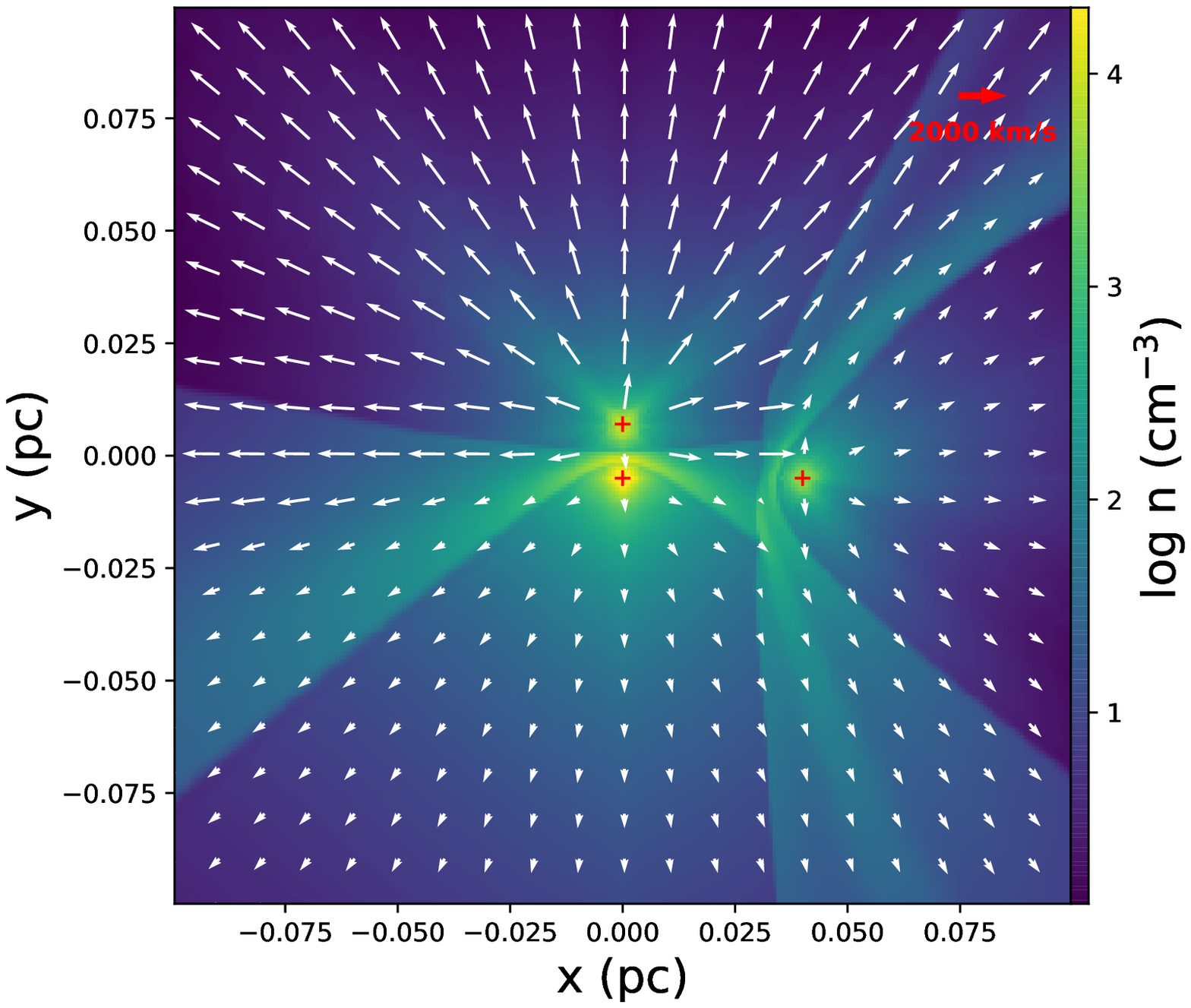}
    \includegraphics[width=0.49\textwidth]{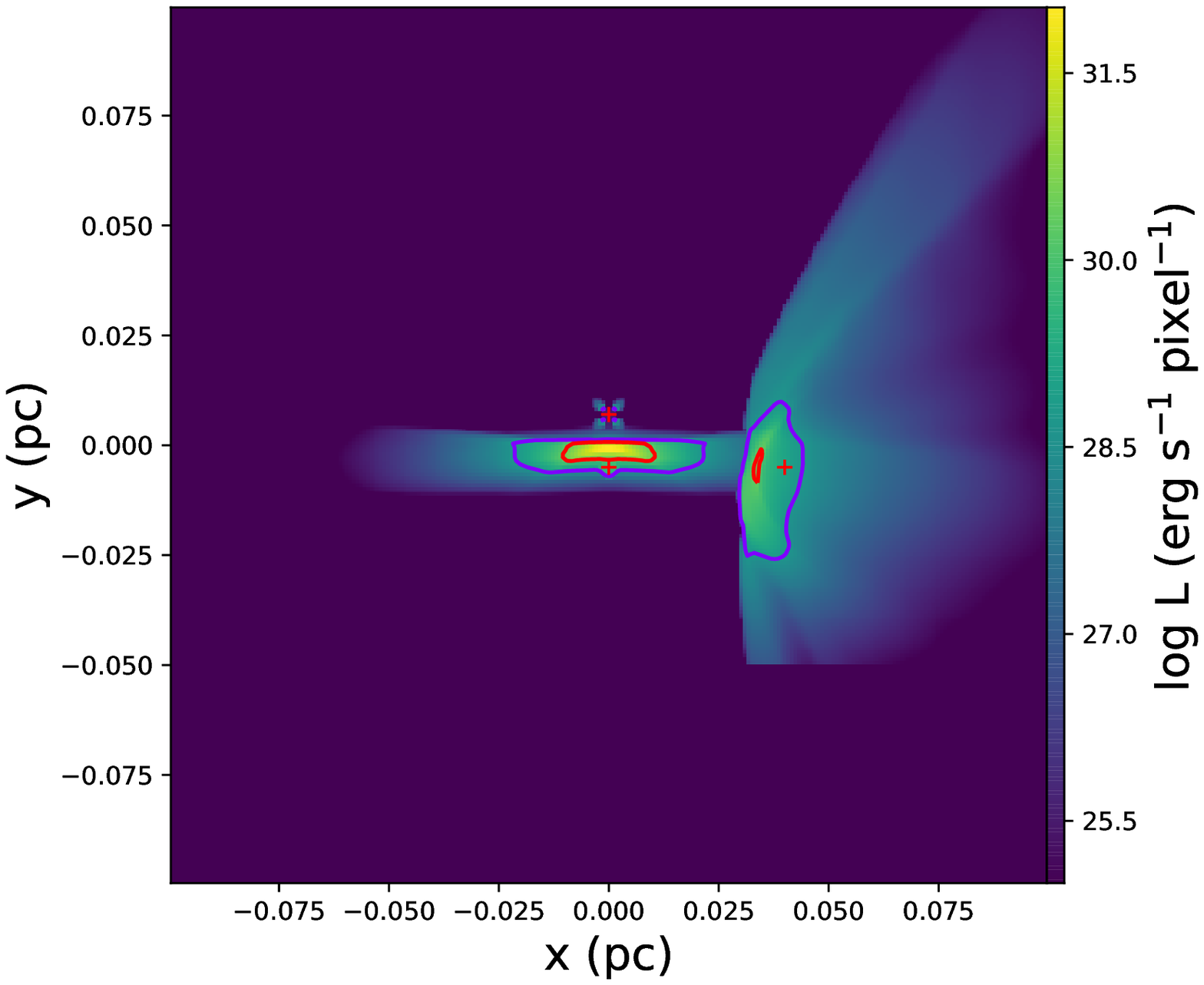}
    \caption{Similar to Figure~\ref{fig:simu}, but a third windy star is added at (x, y, z)=(0.040, -0.005, 0) pc. See Section~\ref{sec:dis} for details. 
      }
\label{fig:simu3}
\end{figure*}

So far we have focused on the wind interaction between the two WR stars. The stellar wind from E1, which is classified as an O supergiant, could have a mutual interaction with the winds of E2 and E4 if it is near enough along the line of sight, and may therefore shape the X-ray morphology.
To test this potential effect, we have run a new set of simulations, adding a third wind source at the projected position of E1 (i.e., assumed to lie also in the $z = 0$ plane). Since there has been no report of the wind mass loss rate or terminal velocity for E1, we adopt $\dot{M}_w = 10^{-6}{\rm~M_\odot~yr^{-1}}$ and $v_w = 1000{\rm~km~s^{-1}}$ which are suitable for O supergiants \citep{2018A&A...620A..89N}. It turns out that E1 has a negligible effect in the X-ray morphology or the total X-ray flux, which is consistent with the observation and can be understood due to its $\sim$10 times lower wind momentum compared to that of E2/E4.
A similar role may be shared by a candidate WR star named E60 (=S4-258 in our {\it Keck} GC source naming convention), located at a projected distance of $\sim$1$\arcsec$ west of E2/E4, originally identified by \citet{Paumard2006}. 
The relation of E60 to IRS13E is unclear, but its proper motion along R.A. has a value of -210 km~s$^{-1}$, quite comparable to the mean value of E1, E2 and E4. 
\citet{Wang2020} suggested that E60 is an interacting binary emitting X-rays and may alone be responsible for the flux associated with what we refer to as the ``tail". 
The X-ray intensity profile examined in Figure~\ref{fig:tail} instead suggests a continuous distribution rather than a sum of two overlapping point sources. 
Hence we consider the scenario that E60 as a third star interacts with E2/E4 and perform a test simulation for this case.
While the wind parameters of E60 are also unknown, we adopt $\dot{M}_w = 10^{-5}{\rm~M_\odot~yr^{-1}}$ and $v_w = 1000{\rm~km~s^{-1}}$, which are again typical of WR stars. 
We find that the wind of E60 is sufficiently strong to affect the X-ray morphology, in the sense that significant X-rays arise from the interacting zone between E60 and IRS13E, as illustrated in Figure~\ref{fig:simu3}. This is consistent with the observed intensity distribution. 
Interestingly, the average temperature of the shock-heated gas in the interaction zone is somewhat higher than that between E2 and E4, thus offering an explanation for the hardness of the tail (Section~\ref{sec:multi}), although this might be a coincidence due to our choice of $v_w$. 
Moreover, the total X-ray luminosity is increased by $\sim$30\%, consistent with the estimated fractional luminosity in the tail. 
The apparent need for E60 to explain the X-ray tail thus suggests that it is located at about the same line-of-sight distance as IRS13E. This raises the possibility that E60 is truly physically associated with the latter. However, the projected offset of E60 from IRS13E is a few times larger than the commonly assumed extent of the cluster, hence E60 is unlikely to be a bound member. 
\\

In the colliding wind geometry, the X-ray centroid should be found near the apex of the colliding wind surface, where gas density is highest and the radiative cooling time is shortest \citep{Stevens1992}. Hence the significant eastward offset of the NIR/radio position of E3 from the X-ray centroid at epoch 2012.0 (Section~\ref{sec:multi}) is rather unexpected, if the former traces the bulk of the strongly compressed and cooled gas and subsequently formed dust \citep{1991MNRAS.252...49U}. 
One possible explanation is that the dusty cool gas, once formed, may drift out from the apex and flow along the colliding wind surface, preferentially to the east, given that the shock caused by E60 would impede flow to the west (otherwise the shocked gas would squirt out in both directions). This is consistent with the slower westward proper motion of E3 compared to both E2 and E4 (Table~\ref{tab:pm}). 
\\

It is worth further comparing the observed radio fluxes of E3 with the expected radio emission from the colliding wind region, presumably dominated by free-free emission from the cool gas \citep{2010MNRAS.403.1633P}.
We collect from the literature flux density measurements of E3, which include: 
VLA 42 GHz (13.10$\pm$1.97 mJy; \citealp{Yusef2014}), ALMA 232 GHz (10.52$\pm$0.90 mJy; \citealp{Tsuboi2019}) and 340 GHz (10.50$\pm$0.50 mJy; \citealp{Tsuboi2017}).
This radio spectral energy distribution (SED), when fitted with a power-law function, $S_\nu \propto \nu^{\alpha}$, exhibits a spectral index $\alpha = -0.1\pm0.2$, fully consistent with free-free emission from a $\sim$$10^4$ K gas \citep{Tsuboi2019}.
The total flux integrated over 40--400 GHz is thus estimated to be $2.5\times10^{32}{\rm~erg~s^{-1}}$.  
If this radio SED is modeled by free-free emission with an electron temperature of $10^4$ K from a spherical volume of $0\farcs1$-diameter, we obtain a gas density of $5.6\times10^5{\rm~cm^{-3}}$.
Extrapolating this SED to the NIR leads to a flux density of 6.2 mJy at 2.2 $\mu$m, $\sim$15 times less than the mean Kp-band flux density (Section~\ref{sec:flux}). This supports the idea that dust emission dominates the NIR flux of E3 \citep{Fritz2010}.
On the other hand, our simulation predicts a 40--400 GHz integrated flux of only $\sim$$10^{30}{\rm~erg~s^{-1}}$ from the colliding wind region.  
3-dimensional simulations of wide colliding-wind binaries, including radiative cooling, also generally find that the radio/millimeter flux is $\lesssim$0.1\% of the X-ray flux \citep{2010MNRAS.403.1633P,2010MNRAS.403.1657P}.
The significance of radiative cooling can be evaluated by the ratio of the cooling and escaping timescales \citep{Stevens1992},
\begin{equation}
\chi = \frac{t_{\rm cool}}{t_{\rm esc}} \approx \frac{(v_w/10^3{\rm~km~s^{-1}})^4 (d/10^{9}{\rm~km})}{\dot{(M}_w/10^{-5}{\rm~M_\odot~yr^{-1}})},
\end{equation}
which is much greater than unity for both the winds of E2 and E4, thus radiative cooling should be insignificant in general.
We speculate that gas cooling and dust formation are episodic, e.g., due to dynamical instability and/or clumpiness of the WR star wind, which have not been self-consistently incorporated in our simulations.
\\

Alternatively, the multi-wavelength emission from E3 may be powered by an embedded IMBH of order $10^4{\rm~M_\odot}$. The {\it Bondi} accretion rate for the putative IMBH follows,
\begin{equation}
\dot{M}_{\rm B} \approx \pi R_{\rm B}^2 v_\infty \rho_\infty = \frac{\pi G^2M_{\rm BH}^2 \rho_\infty}{v_\infty^3}.
\end{equation}
Here the {\it Bondi radius} $R_{\rm B} \approx 0.004$ pc is coincident with the size of E3, given a velocity of the ambient ionized gas $v_\infty \approx 100{\rm~km~s^{-1}}$ as measured from the Br-$\gamma$ line in the Keck/OSIRIS spectrum (Section~\ref{sec:pm}). Thus we obtain $\dot{M}_{\rm B} \approx 4.4\times10^{-5} (M_{\rm BH}/10^4\rm~M_\odot)^2 (n_\infty/5\times10^5{\rm~cm^{-3}})(v_\infty/100{\rm~km~s^{-1}})^{-3} {\rm~M_\odot~yr^{-1}}$. 
Obviously this value is too large to be balanced by the wind injection rate, and in fact becomes a significant fraction of the Eddington accretion rate $\dot{M}_{\rm Edd} \equiv 10 L_{\rm Edd}/c^2 \approx 2.2\times10^{-4}(M_{\rm BH}/10^4\rm~M_\odot){\rm~M_\odot~yr^{-1}}$.
In this case the radiative efficiency of the accretion flow will be $\sim$10\% and thus the expected bolometric luminosity of the putative IMBH would be many orders of magnitude higher than the actual luminosity of E3. 
\\

A much smaller accretion rate may be obtained if the accreted medium is replaced by the X-ray-emitting hot gas. In this case the IMBH is not necessarily embedded in E3, but can find its position somewhere inbetween E2 and E4. 
The gravitational potential of the IMBH is still shallow compared to the kinetic energy of the stellar wind (cf.~definition of the {\it Bondi radius}), hence it will not significantly alter the colliding wind geometry. We have verified this picture with a test simulation by placing a $10^4{\rm~M_\odot}$ point-symmetric gravity near the apex of the colliding wind surface and keeping all parameters of the fiducial simulation unchanged.
Now the {\it Bondi} accretion rate becomes $\dot{M}_{\rm B} \approx 2.6\times10^{-10} (M_{\rm BH}/10^4\rm~M_\odot)^2 (n_\infty/5\times10^3{\rm~cm^{-3}})(v_\infty/500{\rm~km~s^{-1}})^{-3} {\rm~M_\odot~yr^{-1}}$, where the assumed values of $n_\infty$ and $v_\infty$ are appropriate for the keV hot gas near the colliding wind surface. 
In this case $R_{\rm B} \approx 1.7\times10^{-4}$ pc (or $\sim$4 mas in projection) will be unresolved even with ALMA. 
The value of $\dot{M}_{\rm B}$ is about six orders of magnitude lower than the Eddington limit, hence the accretion will be dictated by a radiatively inefficient, advection-dominated accretion flow (ADAF; \citealp{2014ARA&A..52..529Y}).
At this accretion rate, the radiative efficiency is estimated to be $\sim$$10^{-4}$ \citep{2012MNRAS.427.1580X}, thus giving a bolometric luminosity of $\sim$$5\times10^{31}{\rm~erg~s^{-1}}$. 
This is negligibly small compared to the observed X-ray luminosity of IRS13E, and so to the radio luminosity of E3. 
Therefore, radiation from the putative IMBH will be completely masked by the colliding winding-induced emission. In other words, the multi-wavelength emission from E3 and its vicinity provide no direct evidence for an IMBH. 
\\

To date, the strongest argument for E3 embedding an IMBH of order $10^4{\rm~M_\odot}$ comes from the kinematics of ionized gas traced by spatially-resolved H30$\alpha$ line emission \citep{Tsuboi2017,Tsuboi2019}.  
However, we have shown in the above that such an IMBH, inevitably accreting from the same ionized gas, would have produced a luminous X-ray source. 
This calls into question the interpretation of the gas kinematics. 
In fact, the one-dimensional velocity dispersion of the Br-$\gamma$ line from E3, $90\pm10{\rm~km~s^{-1}}$ (Section~\ref{sec:pm}), is quite consistent with the observed velocity width $\sim$$150{\rm~km~s^{-1}}$ of the H30$\alpha$ line. 
This moderately high velocity can be generated from turbulent motions in the colliding wind region without the need for an IMBH. 
The updated stellar kinematics of IRS13E, disfavoring the presence of an IMBH more massive than a few $10^3{\rm~M_\odot}$, supports this view.
\\

Moreover, we note that NIR emission from several hot gaseous clumps has been observed throughout the central parsec of the GC. In particular, several compact objects (the G objects) have been detected orbiting within 1$''$ from Sgr~A* \citep{Gillessen12, Witzel14, Ciurlo20}. It is possible that stellar winds, which can produce E3, are also responsible for producing these objects \citep{Burkert2012}.
However, recent simulations show that clumps formed in stellar wind collisions are extremely light, therefore not supporting the colliding winds being the origin of the G objects \citep{Calderon2016, Calderon2020}. 
It is also unclear how they would survive a close encounter with the SMBH (as G2 did in 2014). Therefore, at present there is no compelling trace for a common origin between E3 and the G objects.\\

We provide some final remarks comparing our work and that of \citet{Wang2020} .
Although our work bears similarities with \citet{Wang2020} in the use of the {\it Chandra} data set and the employment of hydrodynamic simulations (hence not surprisingly, both studies are led to a similar conclusion on the colliding wind origin for the X-ray source IRS13E), there are several important differences between the two studies:
1) Apart from the X-ray analysis, we have also explored kinetic properties via analyzing the NIR images and spectra from 
our long-term program of Keck observations. With the Keck data, we have provided an updated and more precise proper motion measurement for each component in the IRS13E complex, which allows us to place a strong constraint on 
the mass of a putative IMBH.
2) We have performed a multi-wavelength matching of sources, clarifying the position of the X-ray centroid with respect to the NIR and radio source positions. This new  multi-wavelength spatial information sheds light on the colliding wind scenario, as discussed above.
3) We have identified and analyzed the tail-like X-ray morphology and, with the help of hydrodynamic simulations, found a plausible explanation, i.e., interacting stellar wind from a third windy star (E60). This leads to the interesting implication that E60 might be an unbound member of the IRS13E group. Wang et al., instead, considered E60 as an isolated source in projection.
4) We have carried out an analysis of the X-ray and NIR flux variations.
5) We have performed a more detailed X-ray spectral analysis, recognizing that the NEI model is more adequate than the CIE model in characterizing the observed spectra, while Wang et al. only considered the CIE model. The NEI case is consistent with post-shock gas and further supports the colliding wind scenario.

\section{Summary}
\label{sec:sum}
We have studied the Galactic center object IRS13E, a small group of massive stars and the possible site of a hypothetical IMBH, using multi-wavelength observations and numerical simulations and focusing on understanding the properties and origin of its X-ray emission. 
Our main findings are as follows:
\\
\begin{itemize}
\item{The X-ray centroid of IRS13E is found to be consistent with a position about midway between the two WR stars, E2 and E4, whereas it is significantly offset from E3, a prominent source seen in centimeter/millimeter/NIR bands, which has been suggested to host an IMBH.}

\item{The proper motions of the member stars provide a stringent mass constraint on the putative IMBH, $M_{\rm BH} \le 2.2 \times 10^{3} M_{\odot}$, although this limit becomes looser if these stars all have a large line-of-sight distance from the IMBH.} 

\item{During the 18-year X-ray observing period, no significant short-term or long-term variability is found in IRS13E. The light curve of E3 in the Kp-band also indicates no significant intrinsic flux variation.}

\item{The X-ray spectrum of IRS13E can be modeled by a $\sim$ 2 keV plasma moderately deviating from collisional ionization equilibrium.}

\item{Our three-dimensional hydrodynamical simulations of colliding winds between E2 and E4 match well with the observed X-ray spectrum and luminosity, 
and imply no need for an embedded IMBH. The tail-like X-ray feature of IRS13E can be interpreted as the interaction with stellar winds from a nearby third WR star.}
\end{itemize}

We conclude that the observed multi-wavelength properties of IRS13E are well explained by the colliding wind scenario, 
whereas no compelling evidence is found for IRS13E hosting an IMBH more massive than a few $10^3{\rm~M_\odot}$.
A high-definition multi-wavelength view of the interacting stellar winds in IRS13E might be enabled by future X-ray missions such as {\it AXIS} and {\it Lynx}, and in a nearer term, by the {\it James Webb Space Telescope} through the IR band.

\acknowledgments
The W. M. Keck Observatory is operated as a scientific partnership among the California Institute of Technology, the University of California, and the National Aeronautics and Space Administration. The Observatory was made possible by the generous financial support of the W. M. Keck Foundation. 
The authors wish to recognize and acknowledge the very significant cultural role and reverence that the summit of Maunakea has always had within the indigenous Hawaiian community. 
We thank Gunther Witzel, Gaoyuan Zhang and Ping Zhou for helpful discussions.
Z.Z. and Z.L. acknowledge support by the National Key Research and Development Program of China (2017YFA0402703) and National Natural Science Foundation of China (grants 11873028, 11473010).

\newpage

\bibliography{ref}
\bibliographystyle{aasjournal}


\end{document}